# Terahertz Quasi-BIC Metasurfaces for Ultra-Sensitive Biosensing and High-Speed Wireless Communications

Islam I. Abdulaal,[a, #] Abdelrahman W. A. Elsayed,[b, #] Omar A. M. Abdelraouf, [c, *]

[a] Department of Electronics and Communication Engineering, Alexandria University, Alexandria, 21544, Egypt

[b] Department of Electronics and Communications Engineering, the American University in Cairo, New Cairo 11835, Egypt

[c] Institute of Materials Research and Engineering, Agency for Science, Technology, and Research (A*STAR), 2 Fusionopolis Way, #08-03, Innovis, Singapore 138634, Singapore.

[#] These authors contributed equally

* Corresponding author. Email address: Omar_Abdelrahman@a-star.edu.sg

## Abstract

Bound states in the continuum (BICs) have emerged as a revolutionary paradigm in terahertz (THz) photonics, enabling metasurfaces with theoretically infinite quality factors (Q-factors) and unprecedented light-matter control. This review synthesizes a decade of progress in THz-BIC research, tracing the evolution from foundational symmetry-protected designs to application-optimized quasi-BICs. We dissect multipolar origins, topological robustness, and symmetry-breaking strategies underpinning high-Q resonances, alongside computational frameworks for predictive design. The timeline highlights key milestones: early dielectric metasurfaces with high Q-factor, flexible biosensors achieving microgram detection limits, and Kerker-conditioned gas spectrometers reducing path lengths by few orders of magnitude. Emerging frontiers in reconfigurable MEMS-BICs and chiral quantum photonics are critically evaluated. Despite breakthroughs, scalability barriers persist for 6G integration, including nano-fabrication tolerances, material loss trade-offs, and dynamic control gaps. This review establishes BIC metasurfaces as pivotal enablers of compact, high-efficiency THz technologies poised to bridge the gap between fundamental discovery and commercialization of THz-based 6G communication and MedTech.

# 1 Introduction

The emergence of bound states in the continuum (BICs) represents a paradigmatic shift in terahertz photonics, offering unprecedented opportunities to confine electromagnetic energy without radiative losses through ingenious exploitation of interference phenomena and symmetry principles. These non-radiating eigenstates, embedded within the radiation continuum, manifest theoretically infinite quality factors and enable extraordinary light-matter interactions that transcend conventional limitations imposed by diffraction and material losses. The convergence of BIC physics with terahertz (THz) metasurfaces has catalyzed revolutionary advances in high-performance photonic devices, establishing a new frontier for next-generation wireless communication systems, ultra-sensitive biosensors, and compact spectroscopic platforms.[1-7]

BICs constitute a remarkable class of electromagnetic eigenmodes that remain perfectly confined despite existing within a continuum of propagating waves, defying conventional expectations of radiative decay. These exotic states achieve complete suppression of far-field radiation through sophisticated interference mechanisms while maintaining strong near-field confinement, resulting in theoretically infinite quality factors that approach the fundamental limits of electromagnetic confinement. The mathematical foundation of BICs emerges from the complex eigenvalue problem of open optical systems, where specific eigenfrequencies correspond to purely real values, indicating zero radiative loss and infinite lifetime. Contemporary theoretical frameworks describe BICs as singularities in the complex frequency plane, where the imaginary component vanishes precisely at the BIC condition, leading to divergent Q-factors and enhanced field localization[8-14]. The rich taxonomy of BIC phenomena encompasses three fundamental mechanisms, each exploiting distinct physical principles to achieve radiation suppression. *Symmetry-protected BICs* arise from fundamental incompatibility between the spatial symmetry of confined modes and the radiation channels available for coupling to the far-field. These robust states persist even under moderate structural perturbations, provided the underlying protective symmetry remains intact, making them exceptionally promising for practical device applications.[15] *Friedrich-Wintgen BICs* emerge through destructive interference between multiple resonant pathways, creating conditions where radiative coupling to specific channels is completely suppressed. This mechanism, originally discovered in atomic physics, manifests when two or more modes of identical symmetry interact through a common radiation channel, producing one bright super radiant mode and one dark non-radiative BIC. The exquisite sensitivity of F-W BICs to parameter variations enables precise spectral tuning and offers unique opportunities for reconfigurable photonic systems.[16] *Accidental BICs* occur at isolated points in parameter space where radiative losses vanish due to specific geometric configurations, independent of symmetry considerations. These states exhibit exceptional field enhancement and can be accessed through continuous parameter tuning, providing flexibility for device optimization while maintaining high Q-factors. Recent advances have revealed the topological nature of accidental BICs, characterized by quantized topological charges that govern their robustness against perturbations.[17]

The terahertz frequency range, spanning 0.1-10 THz, presents unique challenges that have historically limited the development of efficient photonic devices and communication systems. Atmospheric absorption by water vapor molecules creates severe propagation losses, particularly at specific resonance frequencies, constraining the effective transmission range and limiting practical applications. The absorption coefficient of water vapor reaches approximately 250 $cm^{-1}$ at 1 THz, necessitating innovative approaches to overcome these fundamental limitations. Diffraction limitations become increasingly problematic at THz frequencies, where the wavelength-scale dimensions of conventional optical components lead to substantial beam divergence and reduced focusing efficiency. Traditional approaches using bulk optical elements suffer from size constraints and mounting complexity that impede device miniaturization and system integration. The free space path loss increases dramatically with frequency, following Friis transmission equation, requiring sophisticated beam steering and focusing technologies to maintain adequate signal-to-noise ratios over practical distances.[18-20]

BIC-enabled metasurfaces offer transformative solutions to these challenges through extreme field confinement that dramatically enhances light-matter interactions within subwavelength volumes. The infinite Q-factors associated with ideal BICs enable electromagnetic energy to be trapped and concentrated with unprecedented efficiency, overcoming diffraction limits and enabling device dimensions far below conventional constraints. Loss suppression mechanisms inherent to BIC physics provide pathways to circumvent atmospheric absorption through enhanced coupling efficiency and reduced interaction volumes. The integration of BIC physics with metasurface platforms enables revolutionary flat optics paradigms that replace conventional bulky components with ultra-thin, planar devices exhibiting superior performance characteristics. BIC metasurfaces achieve complete phase and amplitude control across subwavelength-thin profiles, enabling metalenses with diffraction-limited focusing performance and minimal chromatic aberration. Dynamic modulators based on quasi-BIC resonances provide unprecedented modulation depths approaching unity while maintaining compact form factors compatible with integrated photonic systems. The planar architecture of BIC metasurfaces facilitates seamless integration with semiconductor fabrication processes, enabling mass production of high-performance THz components with precise dimensional control and excellent reproducibility.[21-23] Recent advances in reconfigurable intelligent surfaces leverage BIC principles to achieve dynamic beam steering and wavefront shaping capabilities essential for next-generation wireless communication systems (Figure 1).

This comprehensive review addresses the rapidly evolving landscape of terahertz BIC metasurfaces, with particular emphasis on device architectures optimized for THz frequencies and their integration pathways toward practical applications. We examine 6G wireless communication requirements that drive the development of ultra-massive MIMO systems, intelligent reflecting surfaces, and compact THz transceivers capable of supporting unprecedented data rates and connectivity densities. The review critically evaluates translational challenges including fabrication tolerances, material limitations, thermal stability, and system-level integration complexities that currently impede widespread commercialization. Our analysis encompasses recent breakthrough demonstrations in biosensing applications, where BIC-enhanced sensitivity enables detection limits approaching single-molecule resolution for medical diagnostics and environmental monitoring. We further explore emerging applications in nonlinear optics, where BIC states offer unique opportunities for harmonic generation, quantum state manipulation, and coherent light-matter interfaces. The roadmap toward practical deployment addresses fundamental questions of challenges, scalability, robustness, and cost-effectiveness that will ultimately determine the success of BIC technologies in commercial applications.

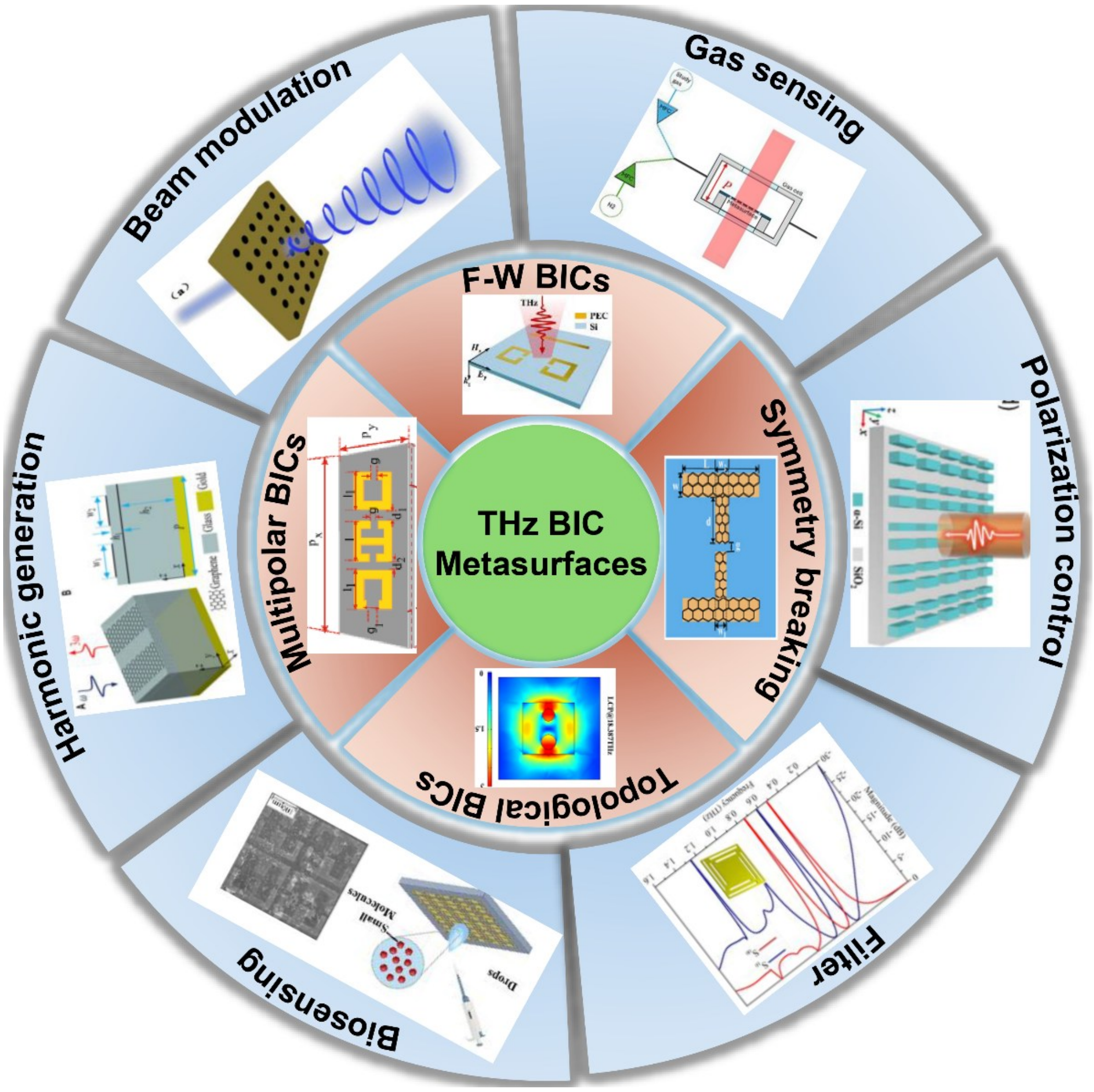

**Figure 1:** Overview of terahertz-based metasurfaces supporting different types of bound states in the continuum resonances and developed applications. Reprinted with permission from.[24-33]

# 2 Theoretical Foundations of BICs in Terahertz Regime

## 2.1 Multipolar Origins

The electromagnetic response of THz BIC metasurfaces originates from sophisticated multipolar excitations that extend beyond conventional electric and magnetic dipole moments to encompass exotic toroidal dipole configurations. Toroidal dipoles arise from closed-loop arrangements of magnetic dipoles or vortex-like current distributions that produce unique radiation patterns fundamentally different from conventional multipolar sources. The mathematical description of toroidal moments involves circulation of electric or magnetic dipole moments around a closed contour, creating electromagnetic field configurations that couple weakly to far-field radiation while maintaining strong near-field confinement.[34] Magnetic toroidal dipoles (MTD) emerge from head-to-tail arrangements of magnetic dipole moments forming toroidal current loops, while electric toroidal dipoles (ETD) result from closed circulation of electric dipole moments. These higher-order multipoles exhibit enhanced Q-factors reaching $10^6$ or higher due to their reduced radiation efficiency compared to conventional dipoles. Recent theoretical advances have identified magnetic toroidal quadrupoles and electric toroidal quadrupoles as even higher-order excitations that provide further enhancement of field confinement and Q-factor performance.[35, 36]

Fano resonances provide the spectroscopic signature of BIC physics through characteristic asymmetric lineshape arising from interference between bright and dark modes. The temporal coupled-mode theory describes Fano profiles through the canonical expression:

$$T(\epsilon) = T_0 \frac{(q+\varepsilon)^2}{1+\varepsilon^2} \tag{1}$$

where $T(\epsilon)$ is the transmission at normalized energy, $T_0$ is background transmission, $\varepsilon = \frac{\omega - \omega_0}{\gamma}$ represents the normalized detuning, $q$ denotes the Fano asymmetry parameter, and $\gamma$ characterizes the resonance linewidth. The Fano asymmetry parameter $q$ encodes the relative strength of coupling between discrete bound states and the radiation continuum, with $|q| \to \infty$ indicating approach to the ideal BIC condition.[37]

Friedrich-Wintgen BICs manifest through destructive interference between multiple resonant pathways coupling to identical radiation channels. The formation mechanism involves two interacting modes of the same symmetry that hybridize to produce one super radiant bright mode and one trapped dark mode. The mathematical framework employs non-Hermitian matrix formalism where the system Hamiltonian includes radiative loss terms:

$$H = \begin{pmatrix} \omega_1 - i\gamma_1 & g \\ g & \omega_2 - i\gamma_2 \end{pmatrix} \tag{2}$$

$\omega_1$ and $\omega_2$ are the resonant frequencies of the two uncoupled modes. $\gamma_1$ and $\gamma_2$ are the radiative loss rates of the two modes, with $i\gamma$ representing the decay due to coupling with the continuum. $g$ is the coupling strength between the two modes. The eigenvalues of this matrix determine the complex frequencies of the hybridized modes, with F-W BIC conditions occurring when one eigenvalue becomes purely real.[38]

The topological protection of BICs emerges from fundamental properties of the complex electromagnetic field distributions in momentum space, characterized by quantized topological charges that ensure exceptional robustness against structural perturbations. Circular polarization singularities ($C$-points) occur where the in-plane electric field components vanish simultaneously, creating phase vortices with integer topological charge $q = \pm 1$. Linear polarization singularities ($V$-points) manifest where the polarization direction becomes undefined due to vanishing field amplitudes. The topological charge of a BIC is defined through the winding of polarization vectors around the singularity point in momentum space:

$$q = \frac{1}{2\pi} \oint \nabla\varphi \, dk \tag{3}$$

where $\varphi$ represents the phase of the complex field amplitude and the integral is evaluated around a closed contour enclosing the singularity. Conservation of topological charge ensures that BICs cannot be destroyed by smooth parameter variations, providing fundamental protection against fabrication imperfections and environmental perturbations.[39, 40] Merging and annihilation of BICs with opposite topological charges create opportunities for dynamic control and enhanced sensitivity. The process follows strict topological selection rules where charges must sum to zero during annihilation events, leading to predictable behavior under parameter tuning. Recent demonstrations have confirmed topological protection in THz metasurfaces through direct observation of polarization vortices using near-field scanning techniques.[17]

## 2.2 Symmetry Breaking for Quasi-BICs

The transition from ideal BICs with infinite Q-factors to practical quasi-BICs with finite but extremely high Q-factors occurs through controlled symmetry breaking that introduces weak coupling to radiation channels while maintaining strong field confinement. The fundamental relationship governing this transition follows the inverse-square scaling law:

$$Q = \frac{Q_0}{\alpha^2} \tag{4}$$

where $\alpha$ represents the asymmetry parameter quantifying the degree of symmetry breaking and $Q_0$ characterizes the intrinsic $Q$-factor of the unperturbed system. Geometric asymmetry parameters include variations in structural dimensions ($\Delta d$), position displacements ($\Delta r$), and shape modifications that break protective symmetries while preserving the essential physics of mode confinement. For displacement-induced asymmetry, the parameter $\alpha$ typically scales as $\alpha = \frac{\Delta r}{a}$, where $a$ represents the characteristic dimension of the unit cell. Size asymmetry follows $\alpha = \frac{\Delta d}{d}$, where $d$ denotes the unperturbed structural dimension.[41]

Parametric optimization of asymmetry enables precise control over quasi-BIC properties, balancing the competing requirements of high $Q$-factor and sufficient radiative coupling for practical applications. Experimental demonstrations in THz metasurfaces have achieved $Q$-factors exceeding $Q = 3700$ through careful control of geometric asymmetries while maintaining robust performance under fabrication tolerances.[15] Angular asymmetry provides an alternative pathway for BIC control through oblique incidence, where the incident angle $\theta$ acts as a continuous tuning parameter. The effective asymmetry parameter becomes $\alpha \approx sin(\theta)$, enabling dynamic tuning of quasi-BIC properties without structural modifications. This approach offers particular advantages for active devices and reconfigurable systems where real-time control over resonance properties is essential (Figure 2).

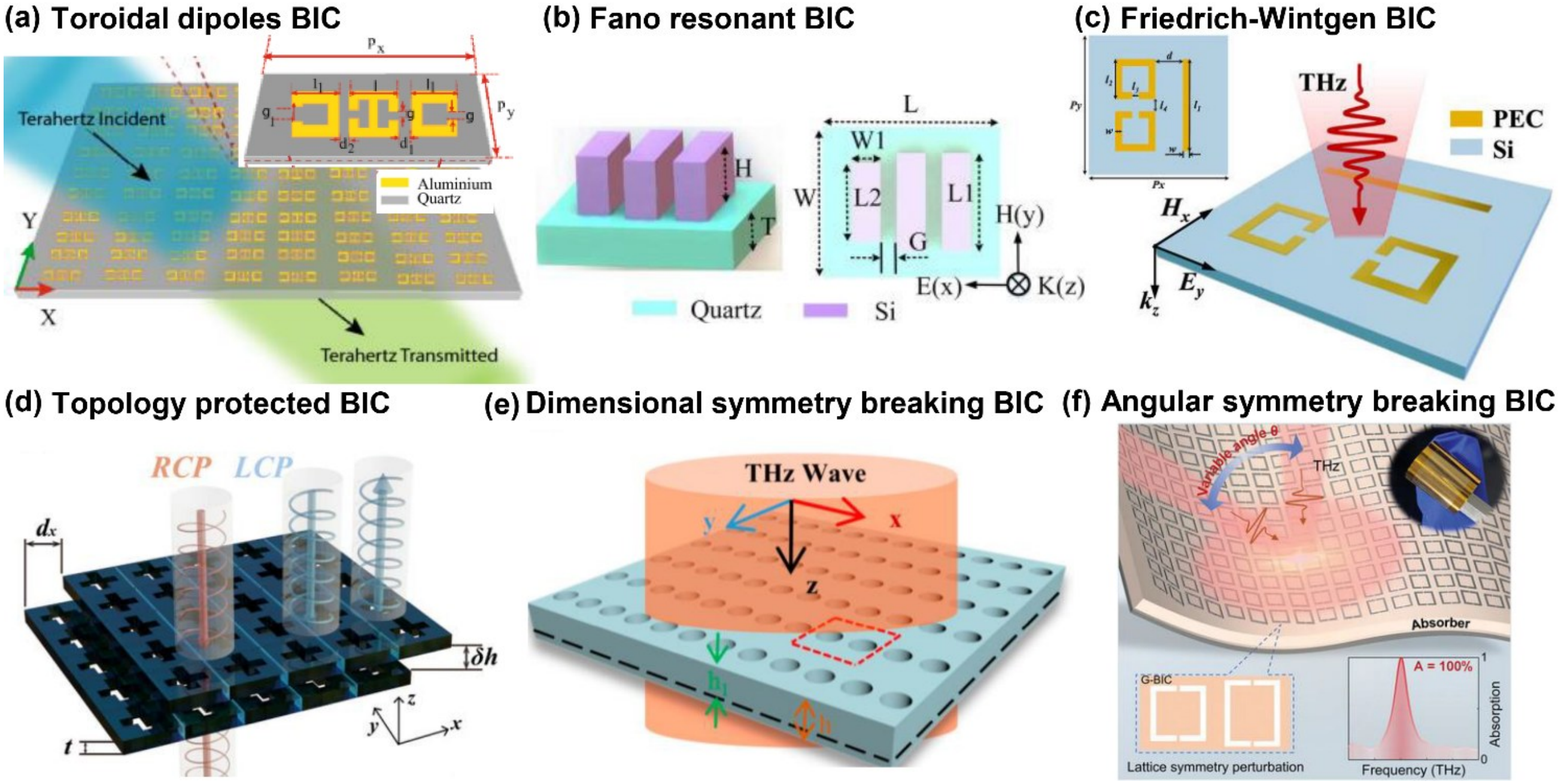

**Figure 2:** Summary of emerging BICs resonances in THz based metasurfaces. (a) Toroidal magnetic dipole BICs. (b) Fano resonant BIC in Si nanopillars. (c) Friedrich-Wintgen BIC in split ring resonators. (d) Topology protected chiral BIC in bilayers metasurface. (e) Dimensional symmetry breaking BIC using different silicon hole radius. (f) Angular symmetry breaking BIC on flexible substrate. Reprinted with permission from.[32, 33, 42-45]

# 3 Timeline of BIC Progress in Terahertz Metasurfaces

## 3.1 Early Demonstrations of BICs

The foundational period of BIC research was marked by pioneering theoretical predictions and first experimental demonstrations of symmetry-protected bound states in all-dielectric metasurfaces. Early work by Gao *et al.* (2016) established the theoretical framework for symmetry-protected BICs, demonstrating that specific point group symmetries could decouple localized modes from radiation channels through fundamental selection rules.[46] The extension of these principles to THz frequencies required sophisticated numerical modeling approaches capable of handling the complex electromagnetic boundary conditions and material dispersion characteristics. First THz BIC demonstrations achieved Q-factors exceeding $10^3$ in all-dielectric silicon metasurfaces operating around 0.5-1.5 THz.[17] These early devices utilized simple geometric motifs including perforated slabs, rod arrays, and disk resonators arranged in periodic lattices with carefully designed symmetry properties. The symmetry-protection mechanism relied on $C_{4v}$ point group symmetry or mirror symmetries that prevented coupling between localized modes and specific radiation channels. Experimental validation employed THz time-domain spectroscopy to measure transmission and reflection spectra, revealing characteristic Fano-like resonances with asymmetric lineshape indicative of BIC physics. Near-field mapping techniques provided direct visualization of electromagnetic field distributions, confirming the predicted anti-symmetric phase relationships between neighboring unit cells that suppress far-field radiation. These measurements represented the first direct experimental observation of BIC field patterns in THz metasurfaces. Material challenges in this early period included ohmic losses in metal components and absorption losses in dielectric materials that limited achievable $Q$-factors below theoretical predictions. The development of low-loss dielectric platforms using high-resistivity silicon and optimized fabrication processes gradually improved performance metrics and enabled more sophisticated device architectures.

The achievement of $Q$-factors exceeding 1000 in all-dielectric THz metasurfaces represented a significant milestone that established the practical viability of BIC-based devices. Silicon-based platforms emerged as the preferred material system due to low intrinsic losses, mature fabrication processes, and excellent mechanical properties suitable for precise dimensional control. Design optimization strategies focused on maximizing the overlap between confined electromagnetic modes and regions of low material loss while minimizing surface roughness and fabrication imperfections that could introduce additional scattering losses. Aspect ratio optimization balanced the competing requirements of strong field confinement and fabrication feasibility. Spectroscopic characterization revealed temperature-dependent $Q$-factor variations arising from thermal expansion effects and temperature-dependent material properties. These observations led to the development of thermally compensated designs and active temperature control systems for high-precision applications.[4]

The development of quasi-BIC resonances in split-ring resonator (SRR) geometries marked a crucial transition from purely academic demonstrations to practical biosensing applications. Classical SRR structures provided an ideal platform for controlled symmetry breaking through gap width variations, angular rotations, or position displacements that could be precisely controlled during fabrication. Metal-based SRR designs initially dominated this development phase due to strong electromagnetic response and well-established fabrication techniques, despite inherent ohmic losses that limited ultimate $Q$-factor performance. The resonance mechanism relied on $LC$ oscillations in the split-ring structure, with magnetic dipole character that could be efficiently coupled to incident THz radiation. Asymmetric SRR configurations achieved quasi-BIC operation through deliberate symmetry breaking using parameters such as unequal gap widths ($\Delta w$), angular misalignment ($\Delta\theta$), or size asymmetry ($\frac{\Delta d}{d}$). The $Q$-factor scaling followed the predicted inverse-square relationship $Q \propto \frac{1}{\alpha^2}$, where $\alpha$ quantified the degree of asymmetry.[47]

## 3.2 THz Metasurface Application Diversification

### 3.2.1 Biosensing:

Biosensing demonstrations utilized the enhanced near-field intensity around quasi-BIC resonances to amplify interactions with biological molecules and tissues. Surface functionalization techniques enabled specific binding of target biomolecules to metasurface regions of maximum field enhancement. Detection schemes monitored spectral shifts in quasi-BIC resonances arising from refractive index changes upon molecular binding. Label-free detection of biomolecules offers many advantages over traditional techniques, which are more prone to error, costly, and often require larger sample sizes.[48] Using THz radiation in label-free detection of biomolecules is suitable due to its unique capacity to probe the vibrational and rotational energy levels of large biomolecules remotely without causing damage, as its low photon energy resonates with these fundamental modes.[49, 50] However, despite their high sensitivity to biomolecular signatures, the practical use of THz waves in biosensing is often overlooked because of the strong absorption of THz radiation by water, which is the main host for most biomolecules.[51] The introduction of quasi-BICs in metasurfaces used for biosensing has fundamentally addressed these limitations by providing a pathway to ultra-high Q-factor resonances, enabling unprecedented light-matter interaction and field confinement.[52]

In 2024, researchers demonstrated a dielectric terahertz metasurface with $C_{4v}$ symmetry that supported a toroidal dipole resonance under normal incidence. By introducing a geometric perturbation, they were able to break the in-plane symmetry, making the setup operative in a quasi-BIC mode. This configuration achieved an ultrahigh Q-factor 12,000, with a refractive index sensitivity of 446 GHz/RIU.[53] In a different paper, Tang et al. demonstrated a practical application of this technology by fabricating and testing an all-dielectric quasi-BIC metasensor based on a hollow-structured silicon design. The sensor was specifically applied to the detection of Auramine O, a carcinogenic dye, achieving a LOD of 0.1 mg/mL and low Q-factor of 30.1, which was a deliberate compromise to ensure detectability with standard terahertz time-domain spectroscopy.[54] Regardless, this marked the potential for this technology in biosensing beyond controlled laboratory settings.

Among advances enabling field deployment of metasurface sensors is the adoption of flexible substrates, which are useful for conformal sensing and enhancing light-matter interactions with metasurfaces. In 2022, Cen et al. demonstrated an early flexible polyimide-based (PI) THz metasurface based on BIC that showed strong near-field enhancement. Further, this biosensing setup developed was cost-effective and easily fabricated.[55] Their conclusion was that substrates with a low refractive index (including polyimide metasurfaces) improve the sensitivity of the biosensing setup.[56] The development of polyimide-based flexible substrates advanced significantly in 2023. Experimental work by Qiu et al. demonstrated a Friedrich-Wintgen (FW) BIC within a polyimide metasurface that presented exceptional mechanical robustness. The quasi-BIC resonance remained spectrally stable even under extreme bending angles up to 180°, and its biosensing capability was validated by detecting Fetal Bovine Serum (FBS) with a low LOD of 0.007 mg/mL[57].

Progress continued into 2025, marked by a diversification of capabilities. Researchers developed a dual-band THz BIC-based metasensor capable of detecting mixtures of Roxithromycin and Bovine Serum Albumin, enabling a LOD of approximately 50 ng/uL.[58] This design featured a high-Q quasi-BIC resonance at ~0.8 THz alongside a broader dipole resonance at ~1.19 THz. Also in 2025, Tiwari et al. examined different properties of metasurface substrates to realize the most optimum design for an effective biosensor.[59] Moreover, Liu et al. made a major leap forward by developing a novel quasi-BIC metasensor integrated with a flexible polyimide capillary. This platform enabled the real-time detection of protein concentrations in aqueous solutions, overcoming the water absorption problem and achieving a remarkable LOD as low as 1 pg/mL for biomolecules such as ovalbumin and hydrolyzed whey protein. This quantification was achieved by monitoring changes in the transmission modulation depth, which exhibited excellent linearity[51].

### 3.2.2 Gas Sensing:

Gas sensing with THz radiation exploits the unique molecular absorption spectra of polar gases, which exhibit strong rotational-vibrational transitions in this band.[60] The usage of metasurface-based sensors in this band leverages the extreme field enhancement and high Q-factors of quasi-BICs to amplify these interactions dramatically, which makes them very attractive for gas sensing and spectroscopy.[61]

In 2022, Zhao et al. devised an early multipolar metasurface, which used a toroidal dipole structure with high-index structure, to reach an ultrasensitive metasurface design that is completely dielectric.[61] Their design supported a toroidal dipole quasi-BIC by breaking the symmetry of the metasurface's unit cells, reaching a theoretical Q-factor of 2,500,000. In addition, this setup was able to achieve high sensitivity, making it a very promising early design for this application. In the following year, Li et al. were able to design a high-Q graphene-dielectric metasurface that supports a multipolar BIC mode. This design broke symmetry through the rotation of two $LiTaO_3$ columns, generating a sensitivity of 0.41 THz/RIU, 0.4 THz/RIU, 1.049 THz/RIU, 0.34 THz/RIU, and 1.59 THz/RIU at peaks 1, 2, 3, 4 and 5 respectively, achieving a quasi-BIC mode. In addition to being multipolar, the designed metasurface could be tuned through the manipulation of the Fermi level of graphene used in the metasurface's structure.[62] This marks a lot of possibilities for applications that could benefit from the adjustment of THz radiation. For instance, in 2025, researchers designed a metasurface that was easier to fabricate and couple with gas-cell architectures in a tunable, and completely dielectric structure. This proposed structure contained a rectangular metasurface with built-in notches and eccentric cylindrical depressions that could control the leakage from the ideal BIC metasurface. This resulted in a Q-factor of 7,211, and a FOM of 951, which despite being low, represent the tradeoff that can be made between potential Q and easier manufacturability and integration potential (Figure 3)[63].

Despite the developments in metasurfaces' tunability and sensitivity through multipolar platforms, there remains some limitations in the light-matter interactions that occur with the metasurface due to the losses induced from the substrate, which reduce the potential Q-factor that could be produced from the setup.[64] One solution to this problem was the implementation of substrate-free metasurfaces, which reduce many of the losses induced in the sensing performance.

In a 2024 paper by Lin et al., researchers designed a metasurface with no substrate in order to achieve more control over the incident light, leading to an almost ideal sensing ability. This experimentally verified setup achieved a high Q-factor, while being in quasi-BIC mode, thanks to the in-plane symmetry breaking in the metallic structure. The setup was able to reach a sensitivity of 0.86 GHz/RIU in the quasi-BIC mode, in addition to being able to sense ultrathin molecules like L-proline, which it was able to detect in a concentration of around 0.87 nmol.[65] In the following year, researchers used the same design approach to design a substrate-free metasurface capable of achieving near-unity transmission. However, instead of using refractive index change to detect gas molecules, this design used resonance frequency shifts with an absorption-based sensing mechanism to improve gas selectivity. Moreover, their design was able to reach a very compact design, with an equivalent path length reduction by 2 to 3 order of magnitude.[60] Further, they were able to design their surface to achieve a Q-factor of around 8,600 for sulfur dioxide, and around 15,000 for hydrogen cyanide with a minimum detection threshold of 10 particles per million (PPM.) These mark a very important milestone in the development of many sensing applications, including real-time gas sensing platforms where the sensor space is limited.

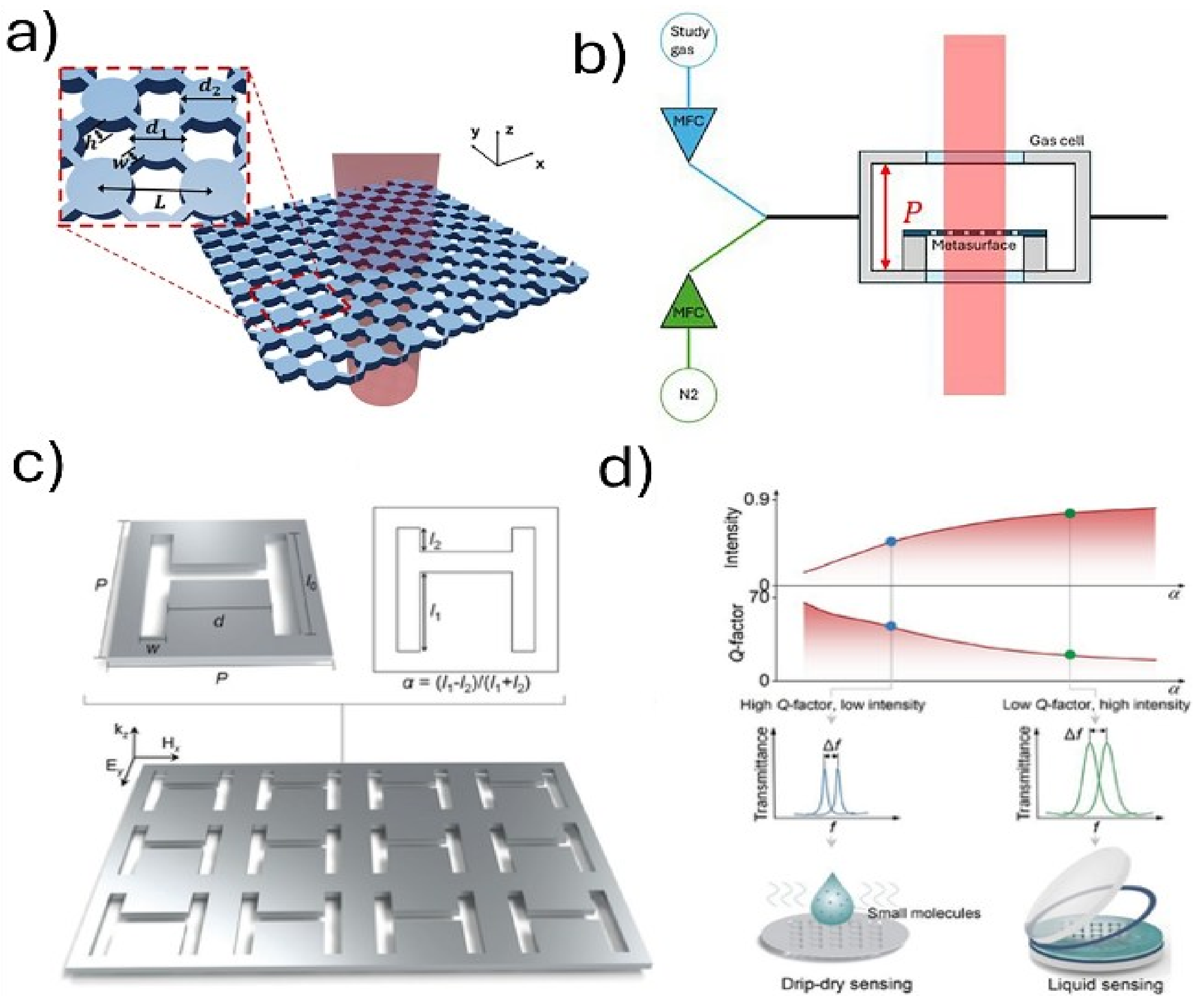

**Figure 3.** Structure of different gas-sensing setups used. a) 3D representation of designed metasurface, showing the unit structure design and dimensions. b) Schematic of the setup used to analyze gas molecules.[24] c) Schematic of metasurface used with geometrical details of the unit cells used in breaking symmetry. d) The effect of the asymmetry factor (α) on the relationship between Q-factor and intensity, where high-Q, low-intensity resonances enable drip-dry molecular sensing, while low-Q, high-intensity modes favor liquid-phase detection through stronger absorption.[66]

### 3.2.3 Polarization Control:

Previous research dedicated to using quasi-BIC metasurfaces operating in the THz regime in the detection of chemical and biological analytes is often bounded by the limitations of symmetry-breaking techniques, which require precise polarization alignment to excite high Q resonances. In real-world sensing scenarios, analytes can depolarize incident radiation during the sensing process, which affects the sensing quality of the setup.[67] In addition, the fact that most THz radiation sources, such as photoconductive antennas, often emit partially polarized, or even unpolarized light, makes the detection process even more difficult.[68] To address this issue, researchers have worked on different techniques that aim to reach polarization-independent metasurfaces that can achieve high resonance factors without being tied to certain polarization alignments (Figure 4).

In 2021, Wang et al. devised a symmetry-breaking metasurface that supports quasi-BIC in the THz range exhibiting polarization-insensitive behavior. This was achieved through a $C_4$ symmetric layout, which is composed of four meta-molecules positioned in a way where each individual meta-molecule is at an orthogonal fashion to adjacent meta-molecules as shown in Figure 4. This induced symmetrical positioning allows for a conserved wave-vector which results in a polarization-insensitive structure, while achieving symmetry-breaking due to the length difference of each individual strip that make up a meta-molecule.[69]

Following up on this, in 2025 a work by Liu et al. was able to demonstrate a polarization-independent quasi-BIC metasurface that uses a unit cell of four diatomic silicon nanodisks arranged in a two-dimensional array that allows for a $C_{4v}$ symmetry profile as shown in Figure 4.[70] This design, which is optimized for near-infrared range, achieves symmetry breaking by changing the nanodisk's radii to be different to each other, exciting a high-Q quasi-BIC while maintaining a rotational symmetry in the structure that provides the polarization-independence. This symmetry profile was achieved in another work by Zhang et al., who was able to develop a graphene-based metasurface that achieves polarization independence in the same wavelength. Their proposed metasurface design uses a monolayer graphene film over a silicon dioxide substrate with four coaxial aperture patterns for every unit cell, as shown in Figure 4. To break the symmetry in the system, the authors change the distances between coaxial apertures labelled as $t_x$ and $t_y$ systematically by the same amount, leading to a quasi-BIC profile without affecting the overall $C_{4v}$ symmetry of the system.[71]

Albeit being very useful in many sensing scenarios, the rotational symmetry induced in the aforementioned techniques can limit the practical applications of these sensors due to having bulky volume cavities and bending radii.[72] To mitigate this issue, Huang et al. had developed metasurface designs that use non-rotational symmetric dimmers to achieve polarization independence, while exhibiting quasi-BIC behavior.[73] Their design was based on the relationship between the magnetic and electric quadrupoles. Researchers have used this relationship to actively tune the asymmetry parameter. They are able to achieve quasi-BICs through using the far-field multipole decomposition and the near-field electromagnetic distribution in their work. Such works mark an important step in the development of metasurfaces supporting gaseous sensing with high quality.

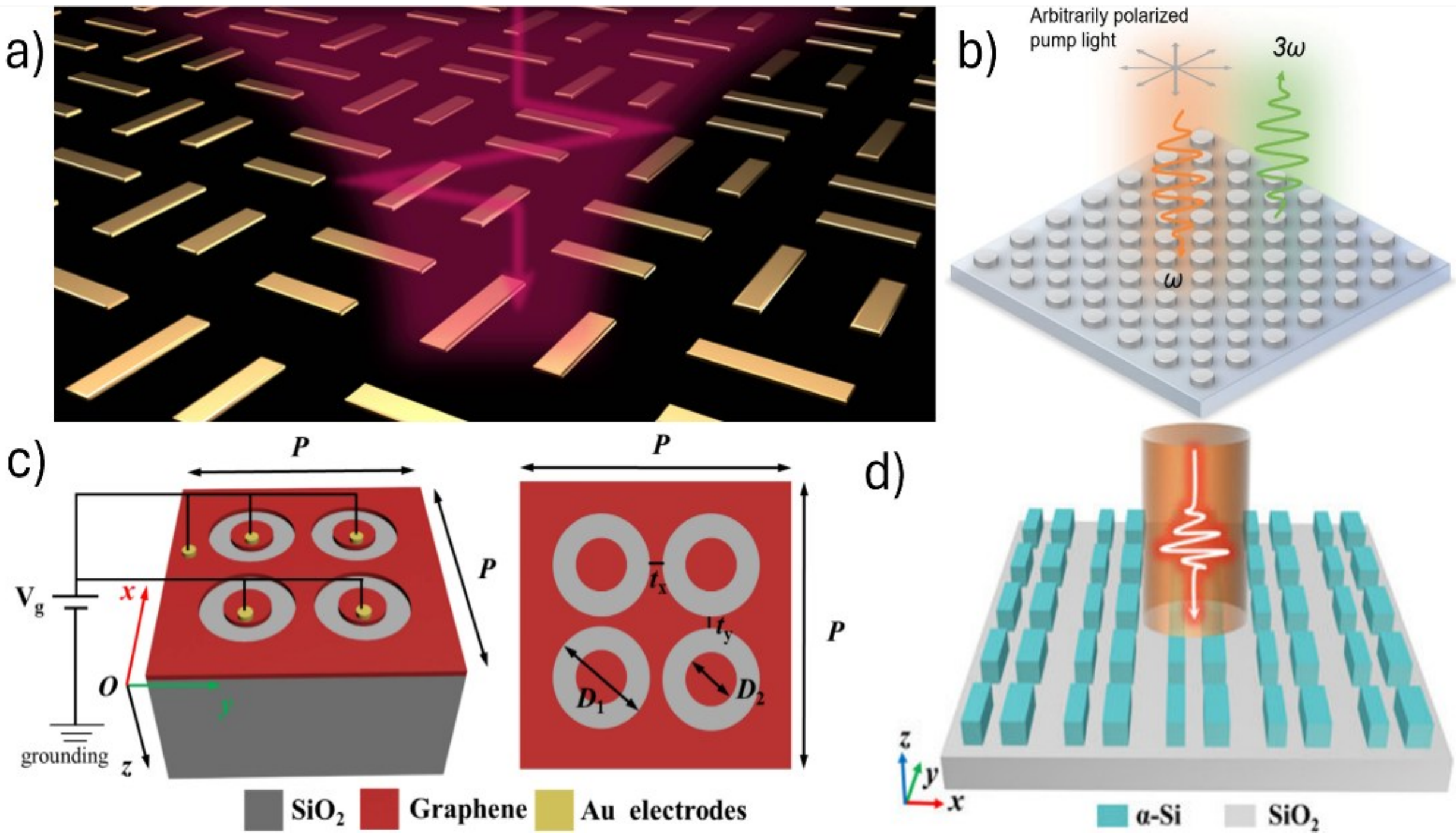

**Figure 4.** Schematics of different techniques used to achieve polarization independence. a) Illustration of THz radiation transmission through the proposed $C_4$-symmetric metasurface.[74] b) Schematic of polarization-independent third-harmonic generation in $C_{4v}$ symmetrical silicon metasurfaces.[75] c) Design structure of dual-band polarization-independent graphene metasurface, with geometric structure of its unit cells.[76] d) Schematic of the nanorod dimer metasurface supporting quasi-BIC resonances for an arbitrary polarized THz incident radiation.[25]

# 4 THz Applications Enabled by BIC Resonances

## 4.1 Ultrasensitive Biosensing

### 4.1.1 Refractive index shifts amplified by linewidth narrowing

One of the major factors that determine the Q-factor of quasi-BIC metasurfaces is linewidth narrowing, which can control the light-matter interactions in biosensing setups. Because of this advantage, researchers have worked on achieving high-Q quasi-BIC setups by optimizing this factor. For instance, Zhang et al. have developed a metasensor exploiting a toroidal dipole quasi-BIC mode to detect amino acids that achieved a remarkably low LOD of 10 pmol/mL. This was achieved by matching the resonance frequency at 0.86 THz to the vibrational fingerprint of the target analyte, which was arginine, and employing a high-Q resonance. The clinical validity of this approach was further confirmed through strong correlation ($R^2 \geq 0.99$) with standard Enzyme Linked Immunosorbent Assay (ELISA) tests on human sweat samples.[77]

Furthermore, symmetry breaking is another major feature that influences the Q-factor achieved by a metasensing setup. A group of researchers demonstrated this by designing a metasurface with two metal open-ring resonators of opposite openings. They introduced a structural asymmetry in the metasurface structure, which transformed a symmetry-protected BIC into a radiating quasi-BIC, yielding a high refractive index sensitivity of 326 GHz/RIU.[78] Further research into the underlying physics reveals that this enhancement is due to a dramatic concentration of electromagnetic energy. For instance, Peng et al. engineered an asymmetry parameter to manipulate the interference coupling between electric quadrupole and magnetic dipole moments.[79] This strategy resulted in a metasurface with a Q-factor of 503 and a refractive index sensitivity of 420 GHz/RIU, accompanied by a 400% increase in confined light energy and a 3300% expansion of the effective sensing area.[79]

Moreover, Wang et al. have reported a metasensor for biodetection with a remarkable sensitivity reaching 674 GHz/RIU.[80] This design, which was based on a split-ring resonator supporting a magnetic dipole quasi-BIC and functionalized with gold nanoparticles, enabled the detection of low-concentration analytes like C-reactive protein (CRP) and Serum Amyloid A (SAA) down to 1 pM. These studies show that strategic symmetry breaking applications based on excited high-Q quasi-BICs are of paramount importance for achieving ultrasensitive biodetection in the THz domain.

#### 4.1.2 Microfluidics-integrated metamaterials with colloidal gold tip enhancement

As mentioned in Section 3.2.1, one of the inherent challenges of the THz wave domain in biosensing is its absorption in aqueous environments. To mitigate this issue, researchers started integrating microfluidics and functionalized nanoparticles to enable precise liquid handling and significant signal amplification. One approach to implement this is the usage of flexible substrates integrated with micro-capillaries, which was demonstrated in 2025 by Liu et al. Their design was able to leverage flexible polyimide capillary effectively to mitigate water absorption and allow for real-time detection of protein concentrations in solutions with an astonishing LOD of 1 pg/mL, which was achieved by monitoring changes in transmission modulation depth (Figure 5) [51].

Moreover, nanoparticles are leveraged for their exceptional field-enhancement properties. For example, Zhang et al. advanced trace liquid detection by integrating colloidal gold onto a quasi-BIC metamaterial. The tip effect of the gold nanoparticles drastically enhanced light-matter interactions, achieving a LOD of 1 ng/mL for imidacloprid solutions.[81] A more sophisticated system was presented by Ma et al., who constructed a hybrid metal-graphene metasurface (MGHM) functionalized with polyethylenimine-modified CuS nanoparticles (PEI@CuS NPs) for bacteremia detection. This platform capitalized on intense electron transfer between the nanoparticles and the graphene layer, which altered the conductivity and induced a measurable shift in the quasi-BIC resonance. This mechanism enabled an ultra-sensitive detection of pathogenic bacteria with a LOD between 11 and 14 Colony-Forming Units (CFU)/mL for different species. In addition, researchers were able to reduce the time-to-positivity (TTP) by an average of 5 hours compared to standard blood cultures, showcasing its direct clinical utility.[82] These developments in ultrasensitive metasensing setups made it easier for more complex analytes like viruses to be detected. For instance, Ding et al. developed a dual-band, all-dielectric metasensor that was capable of detecting with specific viruses such as Hepatitis B (HBV) and H9N2.[83] This device achieved ultra-high Q-factors reaching 1,674 and 2,244. Such advancements in the field of ultrasensitivity showcase the potential of these microfluidic-enhanced lab-on-a-chip devices for the detection of detecting clinically relevant concentrations.

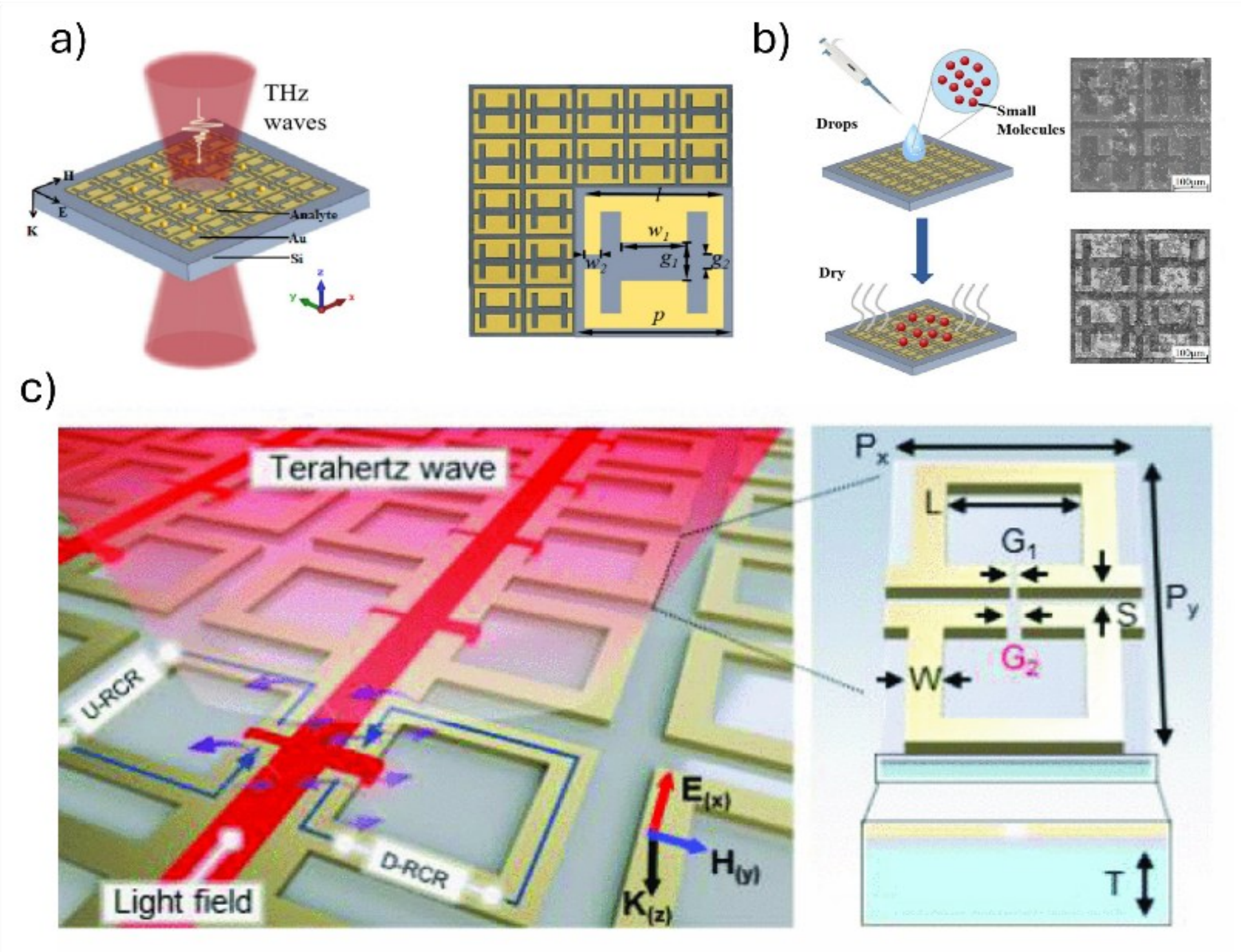

**Figure 5.** Schematic representations of different biosensor setups using microfluidics-integrated metamaterials. a) Detection setup of [26]'s with unit-cell dimensions and THz interrogation and geometry, along with the used substrate delay. b) Sample loading technique demonstrating the concentration-dependent resonance shifts left.[26] c) Structure of [84]'s metallic quasi-BIC unit cell with paired ring-chain resonators and asymmetry converting a symmetry-protected BIC into a narrow quasi-BIC.

### 4.1.3 Specificity-Enhanced Detection: Hybrid Probes and Molecular Recognition

Despite the value of refractive index sensing in many biosensing applications, the ultimate goal of biosensing is to achieve specific molecular recognition. This is accomplished by functionalizing quasi-BIC metasurfaces with biological probes to create hybrid systems that combine physical sensitivity with biochemical specificity. Researchers were able to demonstrate this experimentally through a metasensor that was able to detect CD97, a transmembrane protein specific to gastric cancer cells. These researchers functionalized their asymmetric split-ring resonator metasurface with antibody-conjugated 40 nm gold nanoparticles (AuNPs). The combination of the BIC's field confinement and the high refractive index of the AuNPs allowed for quantitative detection of CD97 across a wide concentration range from 1 pM to 100 nM, presenting a powerful tool for early cancer diagnosis.[85]

The pinnacle of this approach is the ability to distinguish between molecules with high sequence similarity. Shi et al were able to achieve this by designing a system for the specific detection of Let-7a microRNA (miRNA) using a phenylboronic acid-gold nanoprobe (PBA-AuNPs) integrated with a quasi-BIC metasurface. The selective binding of the nanoprobe resulted in a sensitivity for Let-7a that was more than three times greater than for DNA sharing an identical sequence, achieving a remarkable LOD of 0.01 nM.[86] This capability for such precise differentiation demonstrates the potential of quasi-BIC-based platforms for high-fidelity molecular diagnostics. These applications demonstrate the potential of THz-based BIC biosensors as tools capable of complex medical diagnostics. Moreover, scaling light-matter interaction of developed cavities in visible spectrum would allow efficient green energy devices[87-100].

## 4.2 Selective Gas Spectroscopy

In addition to sensing biological samples, current research has diversified into material functionalization and alternative metasurface compositions to target specific gases and enhance practicality. Such efforts are translated into very important applications in multiple areas in gas sensing and detection. For instance, He et al. developed a polarization-insensitive metallic metasensor featuring asymmetric cross-shaped holes to excite dual quasi-BIC resonances. While metallic structures typically suffer from ohmic loss, the quasi-BIC mechanism implemented enabled a high Q-factor and sensitivities of 404.5 GHz/RIU and 578.6 GHz/RIU for the two modes. To achieve selectivity for $CO_2$, the metasurface was functionalized with a polyhexamethylene biguanide (PHMB) polymer layer, showcasing the $CO_2$ detection capabilities with a sensitivity of 0.06 GHz/ppm and a LOD of 33.3 ppm.[101] In another work by with a different setup, researchers utilized the temperature-dependent phase transition of methylammonium lead iodide ($MAPbI_3$) perovskite to create a dynamically reconfigurable quasi-BIC metasensor. The refractive index of the perovskite meta-atoms used could be finely tuned around its 60 °C phase transition, shifting the high-Q quasi-BIC resonance. This platform was applied for $CO_2$ sensing, achieving a maximum sensitivity of 301 GHz/RIU and highlighting a promising route towards adaptable sensors for multi-analyte environments (Figure 6) [102].

Further advancements to improve the potential of metasurfaces in the detection of materials has pushed developments in developments beyond material optimization. For instance, in 2024, researchers advanced the field by designing a metasurface that enables multiple high-Q resonances through the coupling of the building structure of the structure, which included 2 × 2 split ring resonators called metamolecules. As these resonators have two different gap angles, they show two different resonant frequencies, which establish multiple, independent, high-Q resonances on a single metasurface. This design allows for a technique called permittivity retrieval, which works by placing a small, fixed dielectric gratings on some of the resonators, and analyzing the resultant resonance shift from the introduced liquid to the metasurface to get its intrinsic permittivity. This results in effective spectral analysis of the material, instead of just its concentration detection, which was practically verified on glucose by measuring changes in the liquid's properties, and its different concentrations.[103] Such results reveal potentially groundbreaking advancements in the field of sensing.

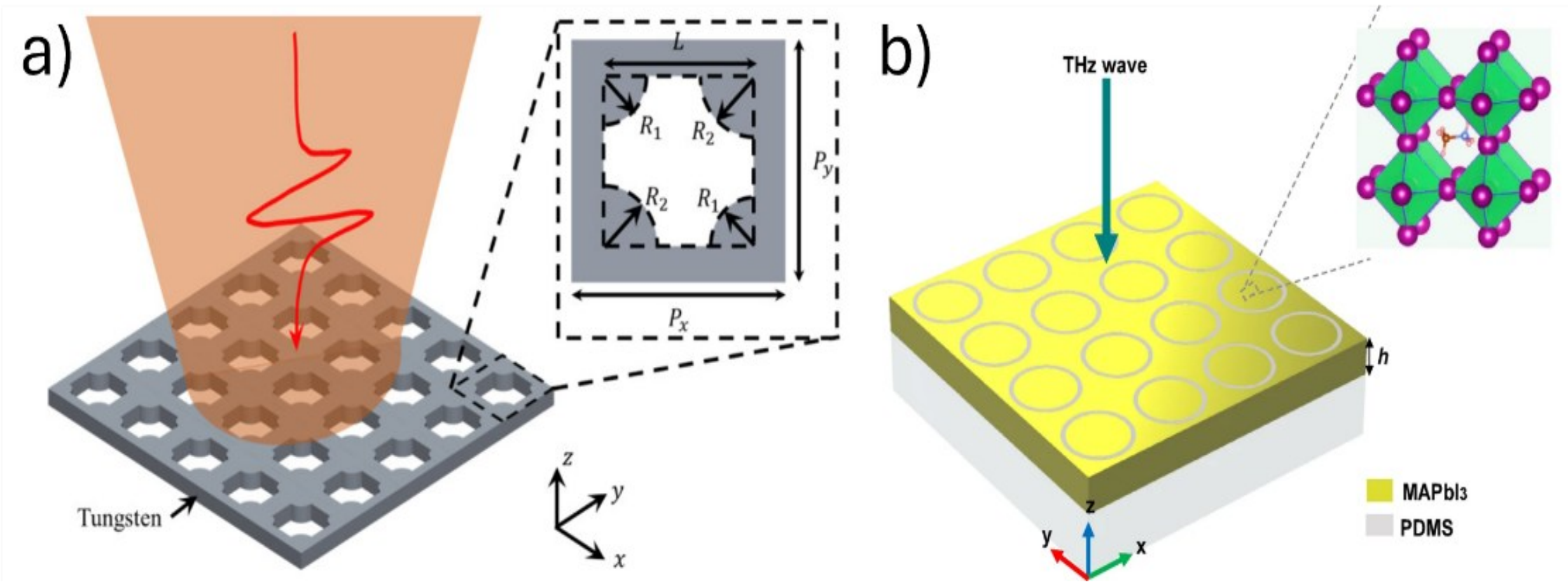

**Figure 6.** Schematics of the metasurfaces used in the detection of detect different gas-sensing setups used. a) Setup structure of the designed metasurface with the asymmetric cross-shaped holes, and the direction of the THz radiation (shown in red)

and covered area (shown in orange).[104] b) Schematic of the metasurface used showing the PDMS and the thin $MAPbI_3$ layer patterned into periodic slot rings, with the direction of the THz radiation.[105]

## 4.3 Polarization/Chiral Control

The development of different designs aimed at improving the polarization insensitivity, and chirality sensing capabilities of metasurfaces have paved the way for more sophisticated applications that were not accessible via traditional metasurfaces. One such application is the development of ultrasensitive refractive and biosensing setups. For instance, researchers were able to exploit polarization-independence to demonstrate a biosensor based on a quasi-BIC metasurface operating in the THz range with low limits of detection. Their proposed sensor was capable of achieving 6.75 THz/RIU with a minimum LOD of 0.0214 RIU. The significance of this design is that it can maintain its quasi-BIC characteristics with various angles of incidence for the light ranging from 0° to 65° regardless of the polarization of the light.[71] In another paper by Yang et al., researchers were able to use a single-crystal silicon chiral metasurface that detects very minute refractive index values, reaching a sensitivity of 1220 THz/RIU for the THz band. Their proposed setup used a phase-detection architecture that amplified the sensitivity of the setup, thanks to the high-Q quasi-BIC design that reached a Q-factor 29800. The designed setup was able to achieve a near-perfect intrinsic and extrinsic chirality with CD values of 0.94, respectively due to the double symmetry breaking, which made the setup intrinsically chiral.[106]

Enantiomer identification, which is the ability to detect chiral molecules, is another area where chiral metasurfaces are used.[107] This application, being important due to its role in the development of many pharmaceutical products, has garnered the attention of many researchers in the field.[108] For instance, Wei et al. successfully used an achiral, pixelated dielectric metasurface to create an integrated sensor capable of simultaneously determining molecular concentration and chiral identity. Their design, which operated via a quasi-BIC mode excited by a novel gap-perturbation technique, generated a strong superchiral field. This enabled an over 160-fold enhancement in vibrational circular dichroism (VCD) signals for the chiral molecule thalidomide, while also providing a theoretical 9 order of magnitude enhancement in absorption signals for concentration measurement.[107]

Moreover, the application of polarization can be extended to other devices, which may benefit from the polarization control within the metasurface setup. For instance, Xu et al. demonstrated how the polarization-dependent quasi-BIC response can be coupled with liquid crystal-driven polarization rotation to give rise to non-reciprocal transmission. This device was able to achieve a tunable isolation of up to 9.8 dB, presenting a significant step toward compact, on-chip THz isolators necessary for robust 6G transceivers.[109] Furthermore, Liu et al. engineered a quasi-BIC metasurface that acts as a highly selective waveplate, capable of converting linear to circular polarization within an ultra-narrow band with an axial ratio (AR) that is less than 3 dB, while exhibiting high sensitivity reaching 6.415 THz/RIU.[110] Such multifunctional applications demonstrate how BIC-enabled field confinement can be harnessed for integrated photonic systems that perform both signal processing and sensing on a single platform.

## 4.4 6G Communication Applications

As illustrated in Figure 7, the integration of bound states in the continuum with terahertz metasurfaces enhances the performance of filtering, beam steering, modulation, switching, multiplexing, and nonlinear applications. This subsection reviews the chronological evolution of communication technologies.

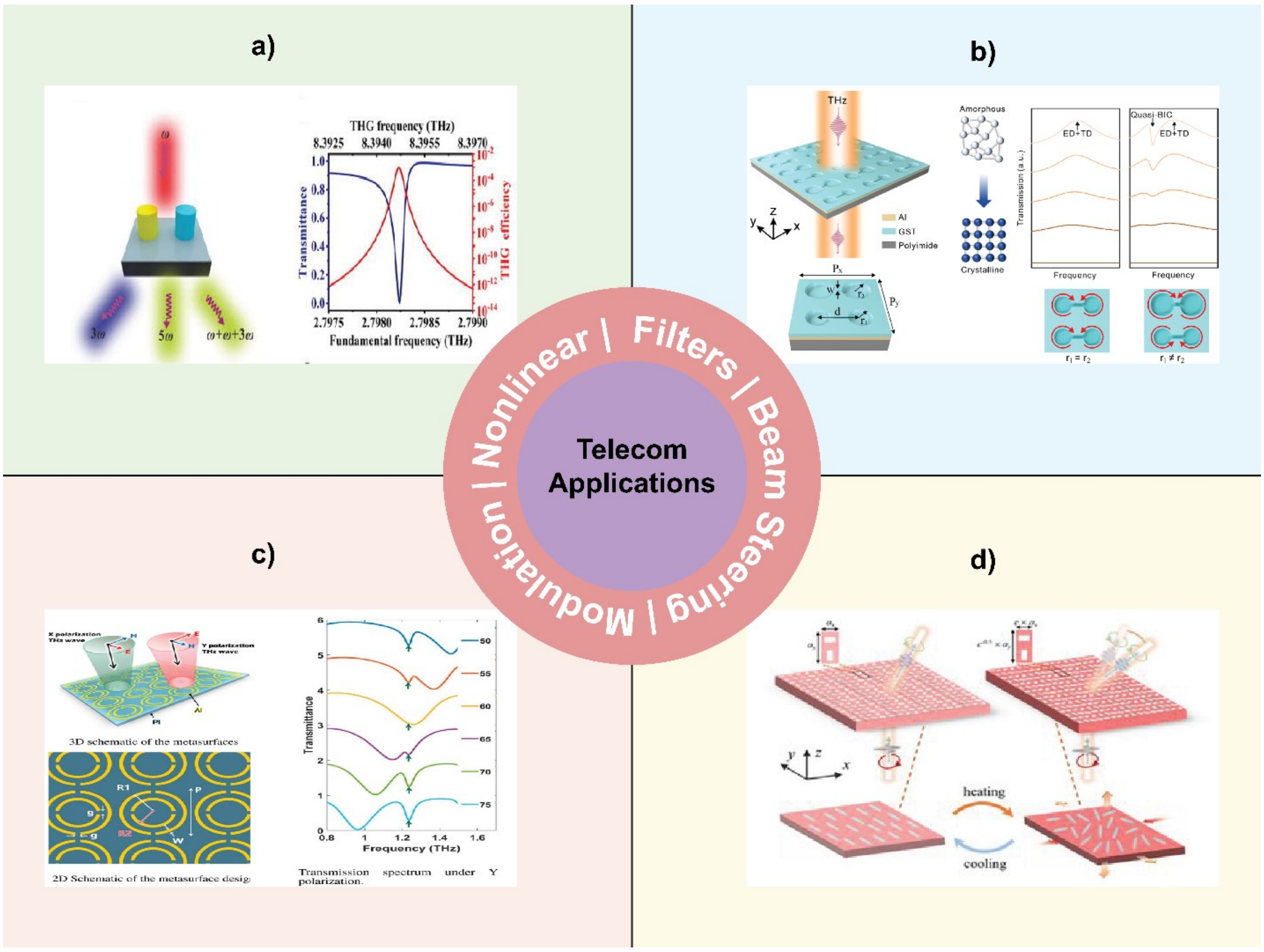

**Figure 7.** Several telecom applications based on BIC, a) Nonlinear generation [111], b) Filtering [112], c) Modulation and multiplexing [113], d) Beam Steering [114]

### 4.4.1 Filters

Filters are critical for signal processing. In 2020, Han et al. established guided-mode resonances in all-dielectric structures, demonstrating periodic silicon cuboid lattices that functioned as diffraction gratings and waveguides while achieving high-Q transmission filters [115]. Asymmetrical designs have emerged. In 2021, Islam et al. established quality factors of 236 using genetic algorithm optimization of elliptical silicon blocks [116]. The period from 2023 to 2024 marks progress toward dual-band systems and the integration of phase-change materials. Perov developed metal mesh filters with dual-band characteristics at 0.516 and 0.734 THz, with bandwidths of 25% and 17%, respectively [117]. Liu et al. demonstrated metasurfaces based on electromagnetically induced transparency effects, achieving peak transmission rates of approximately 0.9 [118]. Feng et al. addressed fabrication tolerances through indirect perturbation methods using asymmetric silicon cuboids while maintaining high-Q Fano resonances [42]. Zhao et al. demonstrated Q-factors of $1.08\times10^4$ with figure-of-merit values reaching $4.8\times10^6$ [119]. Recent developments in 2025 have focused on switchable functionality and ultra-high performance. Zhou et al. demonstrated $Ge_2Sb_2Te_5$-based metamaterials that enabled transitions between transmission and absorption states [120]. Wang et al. achieved tunable control over toroidal dipole and quasi-BIC modes through aluminum dumbbell aperture arrays [112]. Liu et al.

achieved mixed modes by merging quasi-BICs with Wood anomalies [110, 121]. Fu et al. presented dual-bound state architectures with exceptional slow-light effects [122].

### 4.4.2 Beam Steering and Beam Manipulation

The beam-steering capabilities originate from the fundamental concept of vortex beam generation. In 2021, Bai et al. established terahertz vortex beam generators by exploiting the topological properties of BICs in photonic crystal structures [123]. The field progressed in 2023 when Liu et al. developed switchable optical vortex beam generators based on merging bound states and achieved improved quality factors using cross-shaped resonator designs [124]. Active tunability emerged in 2024-2025 through several approaches. Wang et al. contributed symmetry-breaking methods using double-E structures to control the Fano resonance [125]. Long et al. achieved tunable azimuthal deflection of 3° with 4.5% in-plane deformation while maintaining quality factors of 22 and beamwidths of 5° through liquid crystal elastomer integration [114].

### 4.4.3 Active Modulation and Multiplexing

The development of modulation, switching, and multiplexing capabilities represents unified progress from basic demonstrations to multifunctional systems. The timeline began in 2019, when Han et al. established a framework using all-dielectric active terahertz photonics, demonstrating high-Q silicon supercavities that enabled low-power switching at nanosecond timescales [126]. Early material integration in 2020 focused on graphene-based systems. Wang et al. introduced polarization-based multiplexing, who demonstrated multichannel systems based on dual split ring resonator arrays achieving multiple accidental BICs for X-polarization and Y-polarization multiplexing [113]. Multidimensional capabilities were developed by Xie et al., who developed dual-degree-of-freedom multiplexed metasensors that achieved performance across 0.9-1.6 THz bands with signal enhancement factors exceeding 19 dB [127]. The period from 2024 to 2025 marks substantial advances in switching performance. Wang et al. demonstrated vanadium dioxide phase transitions in $LiTaO_3/SiO_2/VO_2$ metamaterials, achieving switching between quasi-BIC states with quality factors of $1.45\times10^5$ and perfect absorption states with modulation depths of up to 0.84 [128]. Wang et al. achieved modulation depths of 98.1% and 99.9% at dual resonance peaks with chemical potential shifts as small as 50 meV [129]. Zhang et al. demonstrated dual-band quasi-BIC systems with tunable transmission ranges of up to 2.3 THz while maintaining modulation depths exceeding 50% [130].

Xu et al. enabled dynamic control of liquid crystal by demonstrating active control using symmetrically broken double-gap split-ring resonators [131]. Ren et al. achieved ultrafast switching with pump thresholds of 192 $\mu J/cm^2$ and cycle times of 7 ps using low-Q plasmonic metasurfaces [132]. Li et al. developed tunable terahertz absorbers, achieving modulation depths up to 99% with insertion losses as low as 0.062 dB [133]. Huang et al. demonstrated all-silicon systems with optical tuning using low fluence excitation [134], whereas Lu et al. explored the cooperative modulation effects in epitaxial vanadium films [135]. Contemporary multiplexing research emphasizes the use of broadband bimodal systems. Yu et al. presented terahertz broadband bimodal multiplexing detection, achieving signal enhancement factors exceeding 10 dB over 1.25-2.20 THz frequency ranges [136]. Potapov reviewed orbital angular momentum multiplexing channels with intelligent metasurfaces for the dynamic manipulation of electromagnetic waves [137].

### 4.4.4 Nonlinear Processing

Nonlinear processing development began with the theoretical foundations laid by Wang and Panoiu, who outlined how optical bound states enhance nonlinear optics through light-matter interaction control [138]. Early experimental work in 2020 by Liu et al. achieved third-harmonic generation enhancement in plasmonic-graphene metasurfaces using hybrid-guided and BIC modes [139]. Hu et al. demonstrated broadband terahertz generation, achieving 17-fold emission enhancement across 0.1-4.5 THz ranges in lithium niobate films [140]. In 2023, the field progressed through various material systems. Wang et al. developed frequency conversion processes achieving 3% conversion efficiency under 50 kW/cm² fundamental incidence in graphene-metal metasurfaces [141]. Ma et al. achieved tunable control through plasmonic quasi-BIC systems, demonstrating extinction ratios of approximately 9 dB and terahertz-level modulation speeds [142].

Recent implementations have achieved substantial enhancements through design approaches and material integration. Sun et al. demonstrated high-harmonic generation from high-Q quasi-BICs through permittivity perturbation, achieving high conversion efficiency for both third-harmonic and fifth-harmonic generation [111]. Shi et al. achieved optical nonlinearity enhancement using epsilon-near-zero metasurfaces, demonstrating nonlinear refractive index coefficients of $1.63\times10^{-12}$ m²/W with response times of 600 fs [110, 143]. Specialized processing capabilities have emerged through advanced design. Liu et al. demonstrated non-reciprocal transmission, achieving an isolation of 34.28 dB at 0.28 MW/cm² incident intensity using asymmetric groove structures [144]. Wang et al. demonstrated third-harmonic generation at critical coupling, achieving a theoretical maximum absorption and 0.3 conversion efficiency under an incident intensity of 50 kW/cm² [145].

# 5 Challenges Toward 6G Integration

## 5.1 Fabrication and Scalability

### 5.1.1 Nano-Precision Requirements: Sub-µm asymmetries (Δd < 50 nm) for Q > $10^5$

The pursuit of ultra-high-quality factors ($Q > 10^5$) in terahertz (THz) metasurfaces via bound states in the continuum (BICs) is fundamentally a nanofabrication challenge. The underlying physics dictates an inverse power-law relationship, ($Q \propto \Delta d$^-n), between the achieved Q-factor and structural asymmetry parameter [146, 147]. As illustrated in Figure 8, this means that realizing a theoretical Q-factor of $10^5$ in a material such as $LiTaO_3$ requires controlling sub-micrometer asymmetries (Δd) to within tolerances of less than a few nanometers. Such nano-precision is non-negotiable for accessing the strong light-matter interactions needed for 6G components (e.g., filters and multiplexers) and ultrasensitive medical technology sensors. In practice, intrinsic material losses (e.g., phonon-polariton absorption in dielectrics) and, more critically, fabrication imperfections currently constrain these limits.

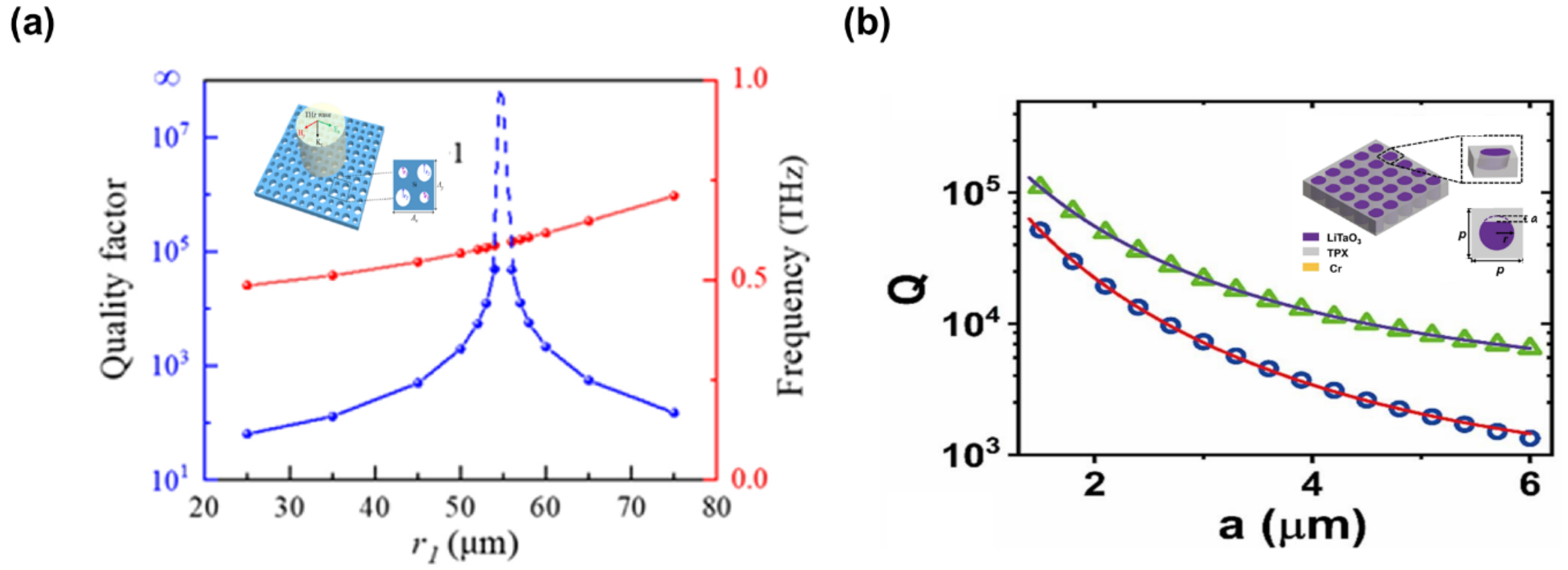

**Figure 8.** a) Q factor across 2D metasurface all silicon with the radius of hole [146], b) Q factor across 3D metamaterial $LiTaO_3$ with the asymmetry parameter (a) [147]

The primary pathway for creating these high-Q structures has been photolithography combined with Deep Reactive Ion Etching (DRIE), which is well suited for the 30 μm–3 mm wavelengths of the THz band (Figure 9) [148]. This approach has yielded some of the highest reported Q factors for all-dielectric BICs to date. For instance, Wang et al. used this method to fabricate an asymmetric silicon circular hole metasurface, achieving a record experimental Q-factor of 1,049 at the time [146]. Similarly, Lin et al. employed dual photolithography and deep silicon etching to create meta-atoms with varying z-axis etch depths, exciting multiple high-Q quasi-BICs (QBICs) for proline and amino acid sensing [149]. A key consideration in this process is achieving a high dielectric contrast to reduce substrate losses, which is a challenge addressed by techniques such as wafer-level anodic bonding of silicon to low-permittivity borosilicate glass [150].

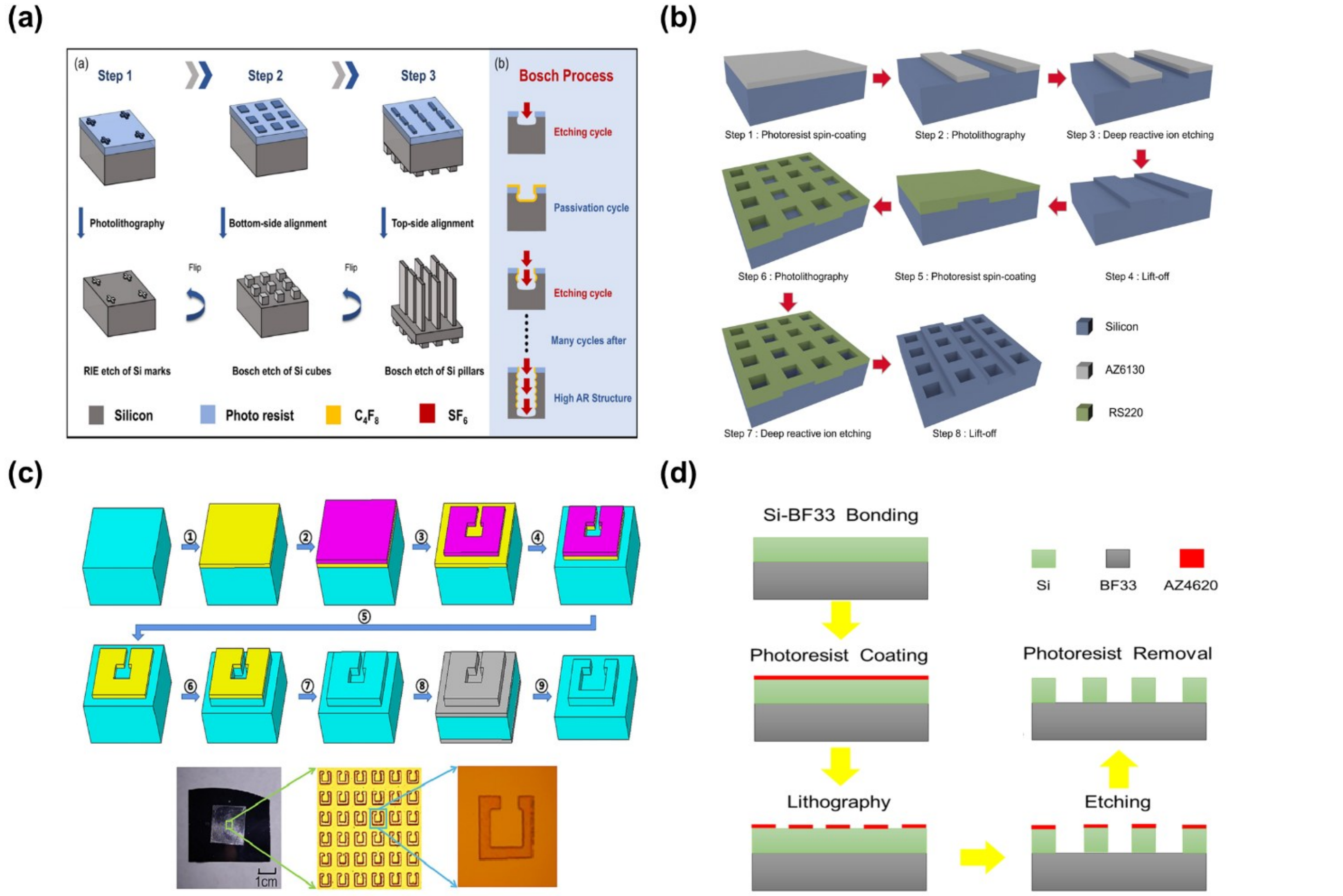

**Figure 9.** Different Photolithography & DRIE Process, a) Bosch process for double-sided silicon metasurfaces [151], b) Dual photolithography for asymmetric silicon metasurfaces [149], c) Metal hard mask etching for high-aspect-ratio structures [152], d) Wafer-scale anodic bonding (Si/glass) [150]

**Table 1.** Comparison of Advanced Fabrication Techniques for High-Q THz BIC Metasurfaces

| Fabrication Technique | Key Characteristics | Primary Materials | Applications for High-Q BICs | Key Challenges |
|---|---|---|---|---|
| Photolithography & DRIE | Resolution ~λ/10, High Aspect Ratio (>12:1) [146, 149] | Silicon (Si) | High. The workhorse for rigid all-dielectric BICs. Apps: High-Q sensors [149, 150], filters. | Limited to 2.5D, sidewall roughness, requires substrate etching. |
| Electron-Beam Lithography (EBL) | Ultra-high Resolution (< 10 nm) [153] | Metals, Dielectrics on various substrates | Critical for R&D. Defines nanoscale asymmetries (Δd) for ultimate Q. Apps: Prototyping complex BIC unit cells, fundamental studies. | Low throughput, serial process, high cost, stitching errors. |
| Nanoimprint Lithography (NIL) | High-throughput, high-resolution (5-10 nm) [154] | Polymers, Resists | High potential for mass production. Apps: Large-area flexible metasurfaces [155], absorbers [156], 6G component arrays. | Mold fabrication, defect density, resist shrinkage. |
| Focused Ion Beam (FIB) Milling | Direct, maskless etching & deposition (~10 nm) [157, 158] | Multilayer stacks | Useful for prototyping. Creates slotted structures and 3D unit cells. Apps: 3D chiral metamaterials [159, 160], complex geometry testing. | Gallium implantation causing losses, serial, slow. |
| Two-Photon Lithography (TPL) | True 3D Nanofabrication (~100 nm resolution) [161] | Photopolymers | Emerging for 3D BICs. Enables volumetric BIC structures (e.g., toroidal moments). Apps: 3D metamaterials [162], complex wavefront control. | Material limitations (typically polymers), often requires secondary processing. |
| Femtosecond Laser Writing (FLDW) | High-precision 3D microstructuring | ITO, Polymers | Good for tunable devices. Apps: Flexible ITO metamaterials for tunable absorption/ modulation [163]. | Resolution limited compared to EBL/FIB. |
| Soft Lithography / Transfer Printing | Patterning on non-planar/fragile substrates [164] | PDMS, Polyimide | Enables flexibility. Transfers pre-fabricated structures. Apps: Flexible metamaterials, conformal sensors. | Alignment challenges, potential for defects during transfer. |
| Self-Assembly | Low-cost, large-area periodic structures [165-167] | Polymer/ceramic composites, Colloids | Low. Limited control over asymmetry. Apps: Primarily for Mie resonators or photonic crystals, not high-Q BICs [165, 166]. | Poor control of nano-asymmetry, defect tolerance, limited material choices. |

The progression of techniques, summarized in Table 1, highlights a central trade-off: the unparalleled precision of research-grade tools such as Electron-Beam Lithography (EBL) is essential for exploring the fundamental limits of the Q-factor, but its serial nature renders it impractical for the wafer-scale production required for 6G [153]. Here, Nanoimprint Lithography (NIL) emerges as a promising high-throughput alternative, capable of replicating master patterns with sub-10 nm resolution over large areas, which is vital for future commercialization and has been demonstrated for flexible, graphene-based metamaterials [154, 155]. Flexibility is paramount in applications requiring conformability or dynamic tuning. Techniques such as Femtosecond Laser Writing have been used to fabricate flexible ITO cross-shaped metamaterials [163], while soft lithography and transfer printing enable the integration of high-quality semiconductor resonators onto flexible PDMS substrates [168]. Looking beyond traditional fabrication, advanced Two-Photon Lithography (TPL) allows for the creation of complex 3D architectures, opening the door to new classes of volumetric BICs that may be more robust to fabrication variances [161].

Concurrently, the exploration of novel materials such as Ruddlesden-Popper 2D perovskites ($PEA_2PbX_4$) offers a path to tunable optical properties and mechanical flexibility, which could relax some alignment and stress-induced precision requirements [169]. These material and topological innovations work in concert with advancing fabrication to push the boundaries, with recent theoretical work demonstrating that carefully engineered out-of-plane asymmetry can yield Q-factors exceeding 6,000 [170]. Navigating the nano-precision requirements for high-Q BICs requires a co-design approach in which the metasurface topology, material selection, and fabrication capability are optimized in unison.

### 5.1.2 Flexibility-Robustness Trade-offs: Polyimide vs. silicon

The tension between mechanical flexibility and optical performance has driven the exploration of alternative substrate materials and fabrication approaches. Traditional silicon-based metasurfaces offer excellent optical properties and fabrication precision but lack mechanical flexibility. Conversely, polymer-based approaches provide flexibility but often sacrifice the Q factor and thermal stability of the device. Self-assembly techniques on flexible substrates have emerged as promising alternatives to these methods. Lan et al. pioneered the integration of high-dielectric constant microspheres ($ZrO_2$) within polydimethylsiloxane (PDMS) matrices, achieving mechanical tunability while maintaining reasonable Q-factors [167]. This approach was further refined by Gao et al., who demonstrated 130 μm average diameter of $ZrO_2$ microspheres (Figure 10)[171].

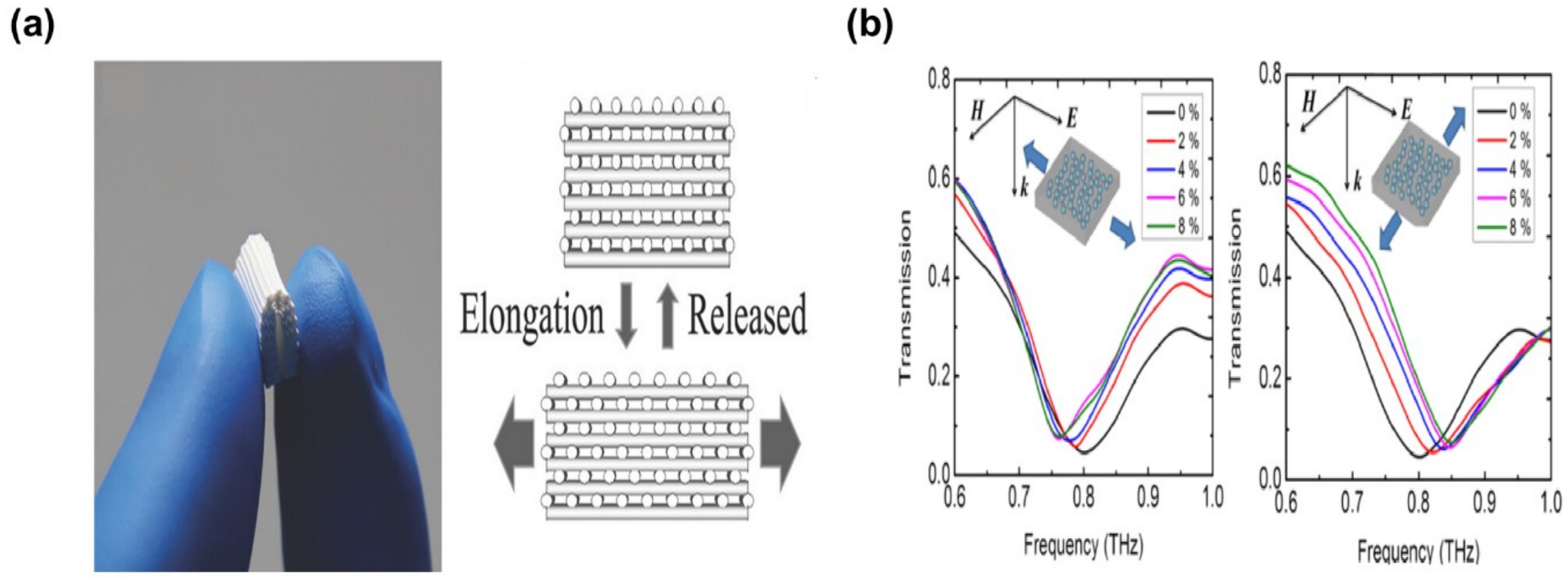

**Figure 10.** Flexibility concept demonstration, a) Flexible material being stretched [172], b) Transmission graphs for different elongation states [167]

Direct writing techniques have provided another pathway for flexible metasurfaces. Zhu et al. developed composite inks combining barium titanate nanoparticles with PDMS, enabling 3D printing of flexible terahertz photonic crystals with tunable properties [172]. By varying the nanoparticle content from 10% to 40%, they adjusted both the refractive index and mechanical properties, achieving a balance between the optical performance and flexibility. Recently, Feng et al. extended this approach using $Al_2O_3$/PDMS composite inks, demonstrating that direct ink writing could produce photonic crystal heterostructures with multiple bandgaps while maintaining mechanical robustness [173]. The trade-off between flexibility and performance remains a fundamental challenge in this area. Yu et al. reported that 3D-printed metasurfaces using acrylonitrile butadiene styrene (ABS) could achieve Q-factors of 2,489 through careful optimization of the printing parameters, although this still falls short of the performance of silicon-based devices [174].

## 5.2 6G-Specific Barriers

Although foundational challenges in fabrication and materials are significant, the integration of THz-BIC systems into the global 6G ecosystem introduces a set of complex, system-level barriers. These are not merely engineering hurdles but fundamental conflicts between new capabilities and existing services and between theoretical device performance and the dynamic control required for functional networks.

### 5.2.1 Interference between communication (active) and Earth Exploration (passive) services

The terahertz (THz) band, a cornerstone of 6G's promised data rates, is not an empty frontier. It is critically occupied by passive Earth Exploration Satellite Services (EESS) and radio astronomy, which detect faint natural THz emissions for climate monitoring, weather forecasting, and atmospheric science [175]. This dual use creates a paramount coexistence challenge, as even low-level emissions from ubiquitous 6G transmitters can irreparably corrupt these exquisitely sensitive passive observations [176, 177]. The interference problem is multifaceted and arises from multiple scenarios, as illustrated in Figure 11. These include emissions from nomadic user devices, reflections from fixed directional links, antenna sidelobes, and airborne THz systems. The core of the problem lies in the extreme sensitivity of EESS receivers, which must detect natural thermal emissions, requiring interference levels to be kept as low as possible [176]. This necessitates stringent equivalent isotropically radiated power (EIRP) limits for terrestrial 6G systems, which directly constrain their range and data capacity.

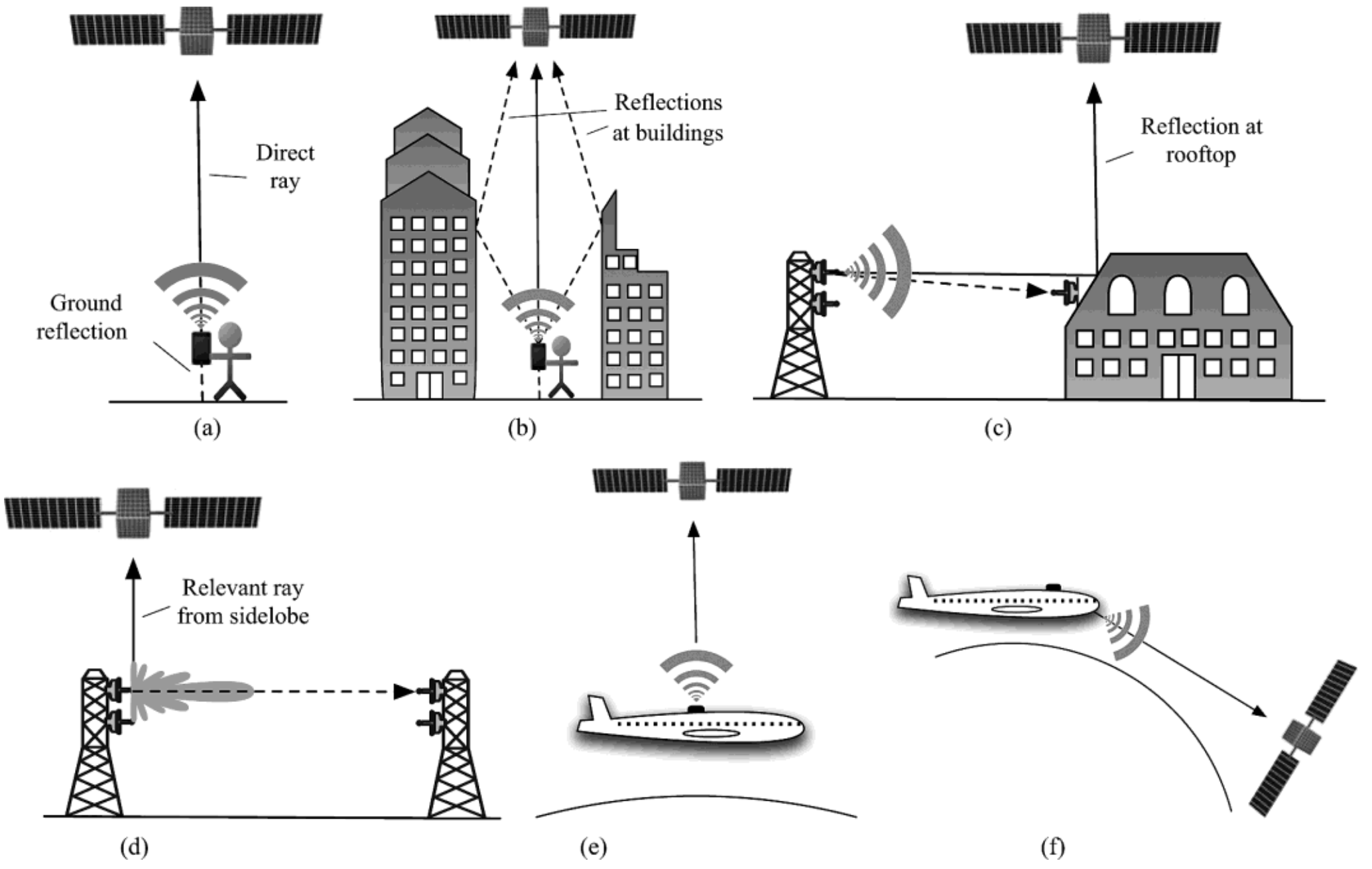

**Figure 11.** shows the possible interference scenarios [176]. (a) Nomadic device in a rural environment. (b) Nomadic devices in an urban environment. (c) Directional link: reflection at the object. (d) Directional link, sidelobe. (e) Airborne THz system. (f) Limb scanning.

Research has progressively quantified these challenges in recent years. Early work by Priebe et al. established strict power spectral density masks for active systems to protect the EESS, highlighting that aggregate interference from dense networks is a primary concern [176]. Subsequent studies have confirmed that large-scale terrestrial deployments introduce significant fundamental interference, requiring sophisticated sharing mechanisms that account for THz-specific propagation characteristics such as molecular absorption and extreme directivity [178, 179]. This problem is further compounded by unintended out-of-band emissions or reflections from reconfigurable intelligent surfaces (RIS), which, while enhancing the signal quality for users, may inadvertently degrade passive sensing [180, 181].

The path forward lies in intelligent and adaptive spectrum sharing methods. Stochastic geometry analyses suggest that with careful network parameter optimization, aggregated interference from heterogeneous networks can be managed below the protection criteria, although this requires a complex system design [177, 182]. Furthermore, the emergence of integrated sensing and communication (ISAC) paradigms offers promising opportunities. Aliaga et al. demonstrated that sub-THz systems can be designed to enhance data rates without substantially degrading the co-located sensing accuracy, although critical trade-offs remain [183]. Ultimately, ensuring harmonious coexistence will require 6G THz systems, including BIC-based RIS, to evolve from being mere transceivers to spectrally aware entities capable of dynamic interference mitigation [180].

### 5.2.2 Dynamic Control: Limited Tunability for Real-time Beam Steering

The unparalleled spectral efficiency of high-Q BIC resonators is paradoxically a significant barrier to their deployment in dynamic 6G networks. The very properties that make them attractive, such as sharp resonance and strong field confinement, make them inherently difficult to tune rapidly. This creates a critical bottleneck in achieving the real-time adaptive beam steering and resource allocation required for 6G applications such as vehicle-to-everything (V2X) communication and UAV coordination [184-186]. This limitation is twofold. First, the switching speed of most tuning mechanisms does not match the 6G requirements. For instance, temperature-based tuning in perovskite quasi-BICs, while achieving high Q-factors, suffers from slow response times associated with thermal diffusion [187]. Similarly, mechanical actuation with liquid crystal elastomers, although offering a good dynamic range, operates on millisecond timescales, which is insufficient for tracking fast-moving mobile terminals [114]. Second, there is a fundamental trade-off between the Q-factor and modulation bandwidth. The long photon lifetime in a high-Q cavity ($Q > 10^5$) physically limits the rate at which the resonance can be shifted or modulated, restricting the system's ability to adapt to channel changes that occur faster than the cavity decay time [135, 188].

Current tuning methodologies each present distinct compromises.

- Graphene-based electrostatic tuning: This method offers faster electronic control but faces challenges in achieving uniform, stable, and precise control over the Fermi level across large arrays, with added complexity in power consumption and biasing circuitry [130, 189, 190].
- Phase-Change Materials (PCMs): Provide non-volatile, bistable switching but are typically limited to binary states, lacking the continuous tunability needed for fine-grained beam steering [114].
- Nonlinear Optical Effects: While BICs can dramatically enhance nonlinearities for frequency conversion, the control complexity of nonlinear systems often precludes the simple, linear control algorithms required for real-time beam management [138].

This "tunability gap" is exacerbated at the system level. The highly directional "pencil beams" necessary to overcome the THz path loss require precise beam alignment. The limited real-time programmability of current THz phased arrays [185, 191] implies that beam management algorithms must contend with hardware that cannot be reconfigured fast enough to maintain seamless handover in high-mobility scenarios. Furthermore, in aerial and space networks, high relative mobility and latency intensify the challenges of Doppler compensation and dynamic beam alignment [192]. Consequently, the development of low-latency, high-speed tuning mechanisms for BIC resonators that can preserve high Q-factors while enabling scale reconfiguration remains one of the most pressing challenges for their 6G integration.

# 6 Comparative analysis and benchmark

The landscape of terahertz BIC-enabled medical technologies has witnessed remarkable evolution from proof-of-concept demonstrations to clinically relevant devices capable of detecting biomarkers at unprecedented sensitivity levels. Biosensing applications represent the most mature commercialization pathway for THz BIC metasurfaces, leveraging their exceptional $Q$-factors exceeding 500 and detection limits reaching single-molecule resolution [15]. Label-free detection paradigms eliminate the need for fluorescent tags or chemical labels, enabling real-time monitoring of biological processes without interference from exogenous markers. Recent breakthrough demonstrations include cardiac troponin I detection at 0.5 pg/mL concentrations, representing an order-of-magnitude improvement over conventional immunoassay. Cancer cell identification through quasi-BIC metasurfaces achieves single-cell sensitivity while maintaining specificity rates exceeding 95% for distinguishing malignant from healthy tissue. Pathogen detection platforms utilizing magnetic nanoparticle separation combined with BIC field enhancement enable SARS-CoV-2 spike protein identification at femtomolar concentrations, far surpassing the sensitivity requirements for early-stage infection diagnosis. The integration of functionalized gold nanoparticles with quasi-BIC resonances provides molecular specificity through antibody-antigen interactions while benefiting from electromagnetic field enhancement factors exceeding $10^3$[193]. Glucose monitoring applications demonstrate the potential for non-invasive diabetes management through skin-penetrating THz radiation that interacts with interstitial fluid glucose concentrations. Broadband frequency-agile metasurfaces achieve glucose detection with molecular-level resolution across the physiologically relevant concentration range (4-20 mM), enabling continuous monitoring without blood sampling.

Terahertz communication systems leveraging BIC metasurfaces address fundamental challenges in 6G wireless networks, including massive bandwidth requirements exceeding 100 Gbps and ultra-low latency constraints below 0.1 milliseconds[194-198]. Ultra-massive MIMO architectures incorporating 1024 or more antenna elements utilize BIC-enhanced beamforming technologies to overcome severe propagation losses inherent to THz frequencies[199]. Hybrid beamforming strategies combine analog RF precoding with digital baseband processing to achieve energy-efficient operation while maintaining high-gain directional transmission. Near-field beam training protocols enable rapid user alignment through multi-beam concurrent scanning, reducing training overhead compared to conventional approaches. The spherical-wave propagation model becomes essential at THz frequencies due to electrically large antenna arrays where far-field assumptions break down within practical communication ranges[200]. Reconfigurable intelligent surfaces (RIS) based on quasi-BIC metasurfaces provide dynamic wavefront control for adaptive channel conditioning and interference mitigation. Beam squint compensation addresses the frequency-dependent beam steering that occurs across multi-GHz THz bandwidths, utilizing true-time-delay architectures to maintain beam integrity. Path loss mitigation through BIC field concentration enables transmission distances extension while maintaining acceptable signal-to-noise ratios for high-data-rate applications. Multi-user beamforming techniques exploit the spatial sparsity of THz channels to serve multiple users simultaneously within the same frequency band. Interference cancellation mechanisms benefit from the highly directional nature of THz propagation, enabling aggressive spatial reuse strategies that multiply spectral efficiency [201].

Quantum photonics applications represent the frontier of BIC research, where enhanced spontaneous parametric down-conversion enables efficient generation of entangled photon pairs with rates exceeding $10^6$ Hz. Chiral metasurfaces supporting quasi-BICs produce circularly polarized entangled photons with unprecedented control over quantum state preparation. The quantum-classical correspondence principle ensures that classical field enhancement directly translates to improved quantum process efficiency. Super chiral electromagnetic fields concentrated within BIC hotspots exhibit helicity densities two orders of magnitude larger than circularly polarized plane waves. These enhanced chiral fields enable enantioselective spectroscopy for distinguishing molecular enantiomers and quantum sensing of chiral biological molecules. Topological protection of BIC states provides robust quantum state generation immune to fabrication imperfections and environmental perturbation [202]. Nonlinear optical applications leverage the extreme field confinement of BICs to achieve frequency conversion processes with dramatically reduced power requirements. Terahertz quantum cascade laser integration with BIC metasurfaces enables coherent emission control and linewidth narrowing through cavity quantum electrodynamics effects. Broadband THz generation through optical rectification in BIC-enhanced nonlinear films spans the frequency range in 0.1–4.5 THz [203].

**Table 2.** Summary of THz metasurface applications

| Application | Material | Q-Factor | Sensitivity (GHz/RIU) | LOD (RIU) | Key Feature | Ref |
|---|---|---|---|---|---|---|
| Refractive Index Sensing | All-Silicon | 3700 | - | - | Ultra-high qBIC | [15] |
| Refractive Index Sensing | Metal Double-Strip | 500 | 87.5 | - | Sharp qBIC | [204] |
| Refractive Index Sensing | Silicon Nanorods | 872 | 512.3 | 197 | Sharp qBIC | [205] |
| Biosensing (Homocysteine) | Silicon 2-Ring Chain | 503 | - | 12.5 pM | Label-free detection | [193] |
| Biosensing (Cancer Cells) | QBIC Metasurface | > 500 | 517 | - | Label-free detection | [206] |
| Temperature Sensing | Silicon Disks | 503 | 569 | $10^{-3}$ | Temperature compensation | [4] |
| Communications (Beamforming) | Hybrid Architecture | - | - | - | Hybrid beamforming | [200] |
| Communications (MIMO) | Ultra-Massive Array | - | - | - | Massive MIMO | [199] |
| Communications (Near-field) | Uniform Circular Array | - | - | - | Multi-beam training | [201] |
| Quantum Photonics | AlGaAs Elliptical | $>10^4$ | - | - | Single photon enhancement | [207] |
| Quantum Photonics | Dual Quasi-BIC | $>10^4$ | - | - | Dual resonance | [202] |
| Chiral Optics | Chiral Dielectric | $>10^3$ | - | - | Circular polarized photons | [208] |

# 7 Outlook and Conclusion

In summary, BICs have evolved from theoretical curiosities to THz technologies, enabling metasurfaces with record high quality factors, ultrasensitive microgram biosensing, and selective gas spectroscopy (ppm-level detection) through symmetry breaking and topological engineering. Despite breakthroughs in flexible platforms and chiral control, scaling for 6G faces critical barriers: nano-fabrication tolerances, dielectric phonon losses, and dynamic tuning gaps for real-time beam steering. To bridge these challenges, future efforts must leverage topology-optimized quasi-BICs, AI-driven design, and quantum-chiral integration, driving toward chip-scale sensors and ultra-massive MIMO beamforming tiles above 100 GHz. Topology optimization frameworks will incorporate fabrication constraints directly into the design process, ensuring that theoretically predicted performance translates to experimentally realizable devices.[209, 210] Real-time design feedback through cloud-based AI platforms will enable researchers to iterate rapidly between simulation, fabrication, and measurement phases. Automated experimental optimization using closed-loop feedback systems adjusts fabrication parameters in real-time to compensate for process variations and achieve target specifications. Multi-fidelity optimization combines low-cost surrogate models with high-accuracy simulations to efficiently navigate complex design spaces. Physics-informed neural networks incorporate Maxwell's equations as constraints during training, will ensure that AI-generated designs satisfy fundamental electromagnetic principles. Transfer learning techniques will enable rapid adaptation of pre-trained models to new materials and frequency ranges, dramatically reducing the data requirements for design optimization. Reinforcement learning agents will explore design spaces through trial-and-error optimization, discovering unconventional architectures that outperform human-designed structures[211, 212]. Integrated quantum circuits incorporating BIC-enhanced sources, on-chip waveguides, and single-photon detectors will enable scalable quantum information processing platforms. Quantum sensing applications will leverage the exceptional field enhancement of BICs to detect single-molecule binding events and sub-femto-tesla magnetic fields.[202] Point-of-care diagnostic devices incorporating BIC biosensors will provide instant pathogen detection for infectious diseases, enabling rapid screening in airports, schools, and healthcare facilities. Multiplexed sensing arrays detect multiple biomarkers simultaneously, providing comprehensive health profiles from single breath or saliva samples. Smartphone integration through miniaturized THz modules will enable distributed health monitoring networks with real-time data analysis and cloud-based diagnostics.[213] Ultra-massive MIMO deployments utilizing BIC metasurfaces will serve hundreds of users simultaneously within single cells while maintaining individual data rates above 10 Gbps. Network slicing through spatially multiplexed beams will provide dedicated resources for autonomous vehicles, augmented reality, and industrial IoT applications. Dynamic spectrum sharing between terrestrial and satellite networks maximizes spectral efficiency across the entire THz band.[1, 7, 214] The final phase addresses spectrum allocation challenges that currently limit commercial deployment of THz communication systems. International Telecommunication Union negotiations require coordinated advocacy to establish global spectrum harmonization above 100 GHz. Coexistence studies with existing services including radio astronomy and atmospheric sensing demonstrate minimal interference potential through advanced beamforming and spatial isolation.[21] By harmonizing photonic innovation with industrial scalability, BIC metasurfaces will ultimately convert the THz gap into a cornerstone of next-generation communication, sensing, and quantum technologies.